\documentclass[superscriptaddress,preprint,amsmath,amssymb]{revtex4-2}

\usepackage{graphicx}
\usepackage{dcolumn}
\usepackage{bm}
\usepackage{lineno}
\usepackage[colorinlistoftodos]{todonotes}

\begin{document}

\title{Spontaneous orbital polarization in the nematic phase of FeSe}

\affiliation{Department of Physics, Massachusetts Institute of Technology, Cambridge, MA 02139, USA}
\affiliation{Advanced Photon Source, Argonne National Laboratory, Lemont, IL 60439, USA}
\affiliation{These authors contributed equally to this work}

\author{Connor A. Occhialini}
\affiliation{Department of Physics, Massachusetts Institute of Technology, Cambridge, MA 02139, USA}
\email{These two authors contributed equally}
\affiliation{These authors contributed equally to this work}

\author{Joshua J. Sanchez}
\affiliation{Department of Physics, Massachusetts Institute of Technology, Cambridge, MA 02139, USA}
\affiliation{These authors contributed equally to this work}

\author{Qian Song}
\affiliation{Department of Physics, Massachusetts Institute of Technology, Cambridge, MA 02139, USA}

\author{Gilberto Fabbris}
\affiliation{Advanced Photon Source, Argonne National Laboratory, Lemont, IL 60439, USA}

\author{Yongseong Choi}
\affiliation{Advanced Photon Source, Argonne National Laboratory, Lemont, IL 60439, USA}

\author{Jong-Woo Kim}
\affiliation{Advanced Photon Source, Argonne National Laboratory, Lemont, IL 60439, USA}

\author{Philip J. Ryan}
\affiliation{Advanced Photon Source, Argonne National Laboratory, Lemont, IL 60439, USA}

\author{Riccardo Comin}
\email{rcomin@mit.edu}
\affiliation{Department of Physics, Massachusetts Institute of Technology, Cambridge, MA 02139, USA}

\date{\today}

\maketitle

\section*{Abstract}

{\bf The origin of nematicity in FeSe remains a critical outstanding question towards understanding unconventional superconductivity in proximity to nematic order. To understand what drives the nematicity, it is essential to determine which electronic degree of freedom admits a spontaneous order parameter independent from the structural distortion. Here, we use X-ray linear dichroism at the Fe K pre-edge to measure the anisotropy of the $3d$ orbital occupation as a function of \textit{\textbf{in situ}} applied stress and temperature across the nematic transition. Along with X-ray diffraction to precisely quantify the strain state, we reveal a lattice-independent, spontaneously-ordered orbital polarization within the nematic phase, as well as an orbital polarizability that diverges as the transition is approached from above. These results provide strong evidence that spontaneous orbital polarization serves as the primary order parameter of the nematic phase.} \\

\clearpage
\section*{Introduction}

Symmetry-breaking phase transitions in strongly correlated electron systems are characterized by their structural and electronic (spin, charge, and orbital) degrees of freedom \cite{Khomskii2014}. In electronically-ordered phases, these degrees of freedom become intertwined, making an experimental determination of the leading interaction challenging. One striking example of this complex interplay is nematicity, an electronically-driven rotational-symmetry breaking which is widely observed in iron-based superconductors \cite{Si2016}. While static nematic order is found to compete with superconductivity in iron pnictide materials such as Co-doped BaFe$_2$As$_2$ \cite{Nandi2010,Pratt2009,Si2016}, it may actually help stabilize it in the chalcogenide FeSe \cite{Liu2018}. Thus, understanding the origin of nematic order in each class is essential for understanding the nature of superconductivity \cite{Chen2020}. Despite sustained efforts, a persistent question is whether nematicity is driven by the spin or orbital degree of freedom, the answer to which has remained elusive due to the complexity of their microscopic relationship to the nematicity-induced structural anisotropy.

The central difficulty in addressing the orbital degree of freedom arises from its close association with the lattice symmetry. This problem is important not just to nematicity in iron-based superconductors, but also to phenomena as diverse as Jahn-Teller distortions in transition-metal oxides \cite{Wilkins2003,Khomskii2014} and quadrupolar $4f$ ordering in heavy fermion materials \cite{Willers2015, Rosenberg2019}. In this work, we introduce a general methodology to distinguish the orbital and lattice degrees of freedom by combining {\it in situ} tunable strain with x-ray linear dichroism (XLD) and x-ray diffraction (XRD), to directly probe the orbital polarization and strain state of the lattice, respectively. We use this methodology to provide the most direct evidence that orbital ordering drives the nematic transition in the iron-based superconductor FeSe.

FeSe displays a nematic transition at $T_s = 90$ K which results in a small lattice orthorhombicity and a large anisotropy in both orbital occupation and spin fluctuations, as demonstrated by many techniques \cite{Li2020,Bohmer2013,Baek2015,Pfau2019,Yi2019,Shimojima2014,Watson2015,Wang2016,Chen2019,Massat2016,Lu2022}. While elastoresistivity measurements have identified an electronic driver for the nematic transition \cite{Hosoi2016,Watson2015,Tanatar2016,Bartlett2021,Ghini2021}, this technique alone cannot uniquely identify the driving interaction \cite{Chinotti2018,Schutt2016}. The primary role of the orbital degree of freedom in the nematic phase has been suggested by nuclear magnetic resonance (NMR) \cite{Baek2015,Li2020} and angle-resolved photoemission spectroscopy (ARPES) \cite{Yi2019, Shimojima2014, Watson2015,Pfau2019} measurements that reveal anisotropy between the Fe $|3d_{xz}\rangle$ and $|3d_{yz}\rangle$ orbitals. Meanwhile, nematicity in the iron pnictide materials is widely thought to result as a vestigial order of antiferromagnetism \cite{Si2016}, and spin-driven nematic fluctuations likely play a major role in the enhancement of superconductivity. However, this spin-nematic model for FeSe is fundamentally challenged by the absence of long-range antiferromagnetic order and by the preservation of nematic order while spin fluctuations are suppressed below the superconducting transition temperature \cite{Baek2015,Bohmer2013}. Intriguingly, these strong spin fluctuations may nonetheless be key to the enhancement of superconductivity while having only a minimal or subdominant role in the formation of nematic order \cite{Baek2020}.

The investigation of nematicity is challenging due to the presence of structural twin domains, which form below the nematic transition and cause bulk probes to average out the electronic anisotropy. Several recent works have used fixed applied strain to (partially or fully) detwin the nematic domains and probe the spin or orbital anisotropy \cite{Huh2020,Lu2022,Chen2019}. However, a fixed strain methodology cannot assess the nature of the coupling between the lattice and electronic degrees of freedom, which is key for understanding what drives the system into a nematic state. Furthermore, any strain applied above the transition induces anisotropy where there is none spontaneously, while within the ordered phase any excess strain applied beyond full detwinning will increase the anisotropy beyond the spontaneous value. Thus, it is necessary to observe the behavior of the electronic anisotropy as a function of both strain and temperature across the nematic transition \cite{Cai2020, Fernandes2013}.

Here, we address this question using strain-dependent XLD at the Fe $K$ pre-edge to probe the local orbital degree of freedom at the Fe site and determine the Fe $3d$ orbital polarization in the nematic state. We use {\it in situ} tunable applied stress to precisely detwin the structural domains and further strain the lattice. These XLD measurements, along with supporting XRD measurements to quantify the strain state, allow us to determine a robust relationship between the orbital and lattice degrees of freedom as a function of strain and temperature across the nematic phase boundary. A key result of this work is the observation of a saturating XLD signal beyond full detwinning in the nematic phase. This suggests a spontaneously-ordered orbital polarization that serves as a primary nematic order parameter, analogous to a saturating magnetization in a ferromagnet. Further, both the strain susceptibility of the orbital polarization (the orbital polarizability) and the simultaneously measured elastoresistivity diverge on approach to the transition from above, consistent with local nematic fluctuations driven by the orbital degree of freedom. While these measurements do not address the spin degree of freedom directly, the transport measurements reveal a secondary source of resistivity anisotropy attributed to anisotropic scattering by spin fluctuations, which is only activated below $T_s$ and is enhanced with increasing orthorhombicity in a strain regime where the orbital polarization is saturated. These measurements thus isolate the critical role of the orbital-degree of freedom in the nematic phase of FeSe and provide direct evidence that nematicity is primarily driven by orbital order.\\

\textbf{XAS and XLD Spectra in Detwinned FeSe}\\

We performed measurements on two samples from the same crystal growth batch. The crystals were cut into rectangular bars with edges parallel to the orthorhombic $\mathbf{a}$/$\mathbf{b}$ directions (Supp. Fig. S1). The samples were then mounted on a titanium support platform as described in Refs. \cite{Park2020, Bartlett2021} and fit with transport leads in a four-wire geometry for determining the longitudinal resistivity $\rho_{xx}$ along the applied stress direction (Fig. 1{\bf b}). Uniaxial stress was applied to the platform with a Razorbill CS130 strain cell, with the nominal strain $\epsilon_{xx}$ determined by the capacitance strain gauge on the stress device (see Methods). 

%We note that as we directly control the total length of the sample via applying stress to the titanium platform, we do not directly observe hysteresis in the detwinning process, in contrast to the methodology involving a sample suspended over a gap as in Ref. \cite{Sanchez2021}.

Sample 1 was initially cooled to $T = 25$ K under moderate tensile stress ($\epsilon_{xx} \simeq 0.1\%$) which partially detwins the structural domains that form below $T_s$. We define the structural A domain (B domain) as that with the longer $\mathbf{a}$ (shorter $\mathbf{b}$) lattice constant aligned parallel to the stress axis (Fig. 1{\bf a}). The inset to Figure 1{\bf c} shows an XRD measurement of the $(114)$ reflection, revealing a 75\% detwinning state, in agreement with the predicted detwinning given the 0.23\% orthorhombicity. Figure 1{\bf c} shows the Fe $K$-edge X-ray absorption spectra (XAS), taken at normal incidence with incident linear polarizations LH/LV (parallel to $\mathbf{a}$/$\mathbf{b}$ in the tensionally detwinned state, respectively) and normalized to the main edge jump (see Methods). The near-edge structure shows three features labeled as A/B/C consistent with previous studies \cite{Lebert2018, Joseph2010,  Simonelli2012}. We report the associated in-plane XLD spectrum in Figure 1{\bf d}, defined as the difference $I_{LV} -I_{LH}$ (for further XLD characterization, see Supp. Figs. S2-S4).

The Fe $K$-edge resonance results from dipole-allowed transitions from Fe $1s$ to Fe $4p$ states, with admixed Fe $3d$ and Se $4p$/$4d$ orbital character due to hybridization. Feature B is the usual main edge of this transition \cite{Joseph2010,Lebert2018}. Higher incident energies (peak C) are dominated by non-local effects including multiple scattering and encode Fe $4p$/Se $4d$ hybridization. Thus, feature C is sensitive to the Fe-Se bond length and long-range structural distortions \cite{Joseph2010, Simonelli2012}. Lower incident energy favors increasingly local electronic states around the absorbing Fe atom \cite{deGroot2009, Modrow2003} and the pre-edge (peak A) coincides with the unoccupied Fe $3d$ density of states near the Fermi level. Due to the local tetrahedral symmetry, the $|3d_{xz}\rangle$ ($|3d_{yz}\rangle$) orbital exhibits strong on-site hybridization with $|4p_{y}\rangle$ ($|4p_{x}\rangle$) states \cite{Fguieredo2022,Chen2012,Yamamoto2008,Lee2009}, giving access to $3d$ orbitals through dipole transitions at the pre-edge (see Supp. Figs. S5-S8). In this picture, the positive pre-edge XLD in the tensionally detwinned state (Fig. 1{\bf d}) corresponds to a more occupied $|3d_{yz}\rangle$ state (Fig. 1{\bf a}), in agreement with the $\Gamma$-point occupation anisotropy seen in ARPES \cite{Yi2019, Pfau2019,Huh2020}.\\

\textbf{Spontaneous Orbital Polarization}\\

We now discuss the simultaneously recorded XLD, XRD and elastoresistivity data collected at fixed strain values on a compressive-to-tensile strain sweep in Sample 1 at $T = 50$ K (see Methods). Figures 2{\bf a-d} show the XLD spectra as a function of strain. The XLD at each XAS feature is strain tunable and reverses sign between compression and tension, consistent with the detwinning of nematic domains. The integrated XLD intensity of the pre-edge peak is plotted versus strain in Figure 2{\bf d}. The XLD increases rapidly before saturating at larger strain, with the saturation occurring near $\epsilon_{xx} \simeq 0.14 \%$. 

On the same strain loop, we used XRD to measure the strain response of the A domain $\mathbf{a}$ and $\mathbf{c}$ lattice constants from which we determine the unidirectional strains $\Delta a/a_0$  and $\Delta c/c_0$ shown in Figure 2{\bf e},{\bf f}. The lattice constants exhibit a weak response to small strain before changing rapidly at large tensile strain, indicating a detwinning strain of $\epsilon_{xx} \simeq 0.14\%$ denoted by red vertical lines in Figure 2. Thus, strains below the detwinning point act to reorient nematic domains with only a negligible effect on the lattice constants of an individual domain \cite{Sanchez2021}. The assignment of the detwinning point is corroborated by the disappearance of the B domain twin peak at the (114) reflection (Supp. Fig. S9-S10). The crossover in the monodomain lattice response is also concomitant with a sign change in the slope of the simultaneously measured resistivity (Fig. 2{\bf g}), discussed in more detail below. 

We thus find that the saturation of the pre-edge XLD coincides with the full detwinning of the sample. In contrast, the higher energy XLD features continue to increase with tension past the detwinning point (Fig.2{\bf b}/{\bf c}, Supp. Fig. S4). This suggests that the higher energy XLD features probe the net lattice anisotropy between the $\hat{x}$ and $\hat{y}$ directions which changes both with detwinning and with strain-enhanced orthorhombicity. Indeed, the structural contributions to the XLD are linear in the orthorhombicity and well-captured by multiple scattering calculations \cite{Chen2010} (see Supp. Figs. S5-S7). From these considerations, we conclude that the saturating pre-edge XLD signature (with integrated value $\Delta n$) corresponds to a lattice-independent, spontaneous polarization of Fe $3d$ orbitals which is only weakly coupled to further structural distortion.  

To further associate the orbital polarization with the emergence of nematicity, we performed XLD measurements on Sample 2 as a function of strain and at fixed temperatures above and below the nematic transition. We plot $\Delta n$ versus strain in Figure 3{\bf c}/{\bf d}. At a given tensile strain value, $\Delta n$ clearly increases with decreasing temperature, while at a given temperature the susceptibility is maximum near zero strain. Using a combination of XRD and optical birefringence measurements (Supp. Figs. S11-S12), we identify the approximate detwinning strain point at each temperature below $T_s$, which coincides with the inflection point of $\Delta n$. These data indicate an orbital polarization that develops with decreasing temperature below $T_s$ with a diminishing strain-susceptibility beyond full detwinning (see Supp. Fig. S13).

Finally, we characterized the orbital polarizability above the nematic transition in Sample 1. To do so, we measured the temperature dependence of $\Delta n$ at a moderate fixed tensile strain of $\epsilon_{xx} \simeq 0.2 \%$ (Figure 4{\bf a}). The constant linear strain state is confirmed by XRD measurements and a fixed orthorhombicity is suggested by the constancy of peak C in the XLD spectrum between 120 K and 90 K, which encodes the structural orthorhombicity (Supp. Fig. S14). Over this same temperature range, the peak A XLD ($\Delta n$) increases strongly (over a factor of 2) before saturating for $T < T_s$. This is quantified with a Curie-Weiss analysis for $T > T_s$, revealing a Curie temperature $T^* = 62.5 \pm 5$ K (Figure 4{\bf a}), thus identifying a divergence of $\Delta n$ as $T_s$ is approached from above. Since this divergence occurs under fixed lattice conditions, we understand $\Delta n$ as originating from a strain alignment of diverging orbital-origin nematic fluctuations and not as a secondary orbital response to the lattice distortion. Below $T_s$, the XLD signal appears to be nearly saturated, even up to the transition itself, which is due to both domain detwinning and additional strain-induced orbital polarization. This demonstrates that a fixed-strain methodology cannot capture the continuous onset of the spontaneous order parameter, which instead requires a fixed-temperature variable-strain approach as in Figs. 2/3.

The combined strain- and temperature-dependence reveals two distinct pieces of evidence regarding the nature of the measured orbital polarization. On one hand, the strain-dependent curves as a function of temperature directly show that a lattice-independent, spontaneously-ordered orbital polarization emerges in the nematic phase (Fig. 3{\bf e} top). At the same time, the strain-susceptibility of the orbital polarization shows a divergence under fixed strain conditions above $T_s$ (Fig. 3{\bf e} bottom). This behavior is reminiscent of the magnetization across a para- to ferromagnetic transition, with the magnetic field being replaced by an anti-symmetric strain as the poling field, and associates the orbital polarization with the primary order parameter of the nematic phase. Therefore, the effect of strain far below the transition is only to reorient `nematic moments’ that are fixed in magnitude and orientationally locked to the underlying structure. Evidence for such a scenario is supported by X-ray pair distribution function experiments \cite{Koch2019}, which associate local nematic moments with a short-range ordered orbital degeneracy lifting that persists far above the nematic transition, consistent with the temperature- and strain-dependence above $T_s$ in our experiments. \\

\textbf{Signatures of Orthorhombicity-Coupled Spin Fluctuations}\\

These conclusions are further supported by simultaneous elastoresistivity measurements, which reveal a close correspondence between the orbital polarization and the resistivity anisotropy above $T_s$ that breaks down within the nematic phase. In Figure 4{\bf b}, we show the temperature-dependent elastoresistance collected simultaneously with the fixed-strain XLD measurements (Fig. 4{\bf a}). Above $T_s$, the resistivity anisotropy ($\Delta \rho /\rho_0$) diverges with a Curie-Weiss temperature dependence to $T^* \simeq 61.4 \pm 3.1$ K, consistent with previous elastoresistivity measurements \cite{Bartlett2021, Ishida2022} and in close correspondence to the Curie temperature of $\Delta n$ ($T^* = 62.5 \pm 5$ K). Below $T_s$, $\Delta n$ saturates while $\Delta \rho /\rho_0$ decreases rapidly, implying a breakdown in the orbital-transport correspondence. 

This is highlighted in more detail in Figure 4{\bf c} (inset), where we plot $\Delta \rho / \rho_0$ vs $\Delta n$ from Fig. 2{\bf d}/{\bf g} for Sample 1 at $T = 50$ K (see Supp. Figs. S15-S18). Here, the resistivity increases linearly with the orbital polarization up to the detwinning point, beyond which it decreases rapidly with increasing strain even while the XLD remains saturated. Equivalent data in Sample 2 show the same orbital-transport linearity up to full detwinning, with a strongly temperature-dependent proportionality (Supp Fig. S16). In Figure 4{\bf c} we plot the slope of $\Delta \rho / \rho_0$ vs $\Delta n$ from both samples across temperature and phase, for both fixed-strain temperature sweeps and fixed-temperature strain sweeps within the detwinning strain regime (see Supp. Figs. S15-S18). We find a positive and weakly-temperature dependent orbital-transport proportionality above $T_s$ which rapidly diminishes below the transition, with a sign change near $T=40$ K. Taken together, these results suggest a second source of elastoresistivity which only becomes dominant below $T_s$ and at strains beyond full detwinning.  

The elastoresistivity encodes information from several distinct but ultimately intertwined sources: orbital polarization primarily creates anisotropy in the Drude weight, while spin fluctuations primarily create anisotropy in the scattering rate \cite{Fanfarillo2016,Onari2017,Fernandes2011,Tanatar2016, Schutt2016,Chinotti2018}. Thus, orbital and spin contributions to the elastoresistivity can behave independently. Our simultaneous XLD and transport measurements enable us to show that the resistivity has a component which closely corresponds to the orbital polarization, both above $T_s$ and within the detwinning strain regime below $T_s$. We propose that the second component of the elastoresistivity originates from spin scattering. FeSe exhibits the same stripe-type spin fluctuations as found in iron pnictide materials \cite{Wang2016,Chen2019,Baek2015,Li2020,Lu2022}, which are thought to drive the large negative elastoresistivity in the latter. In the non-magnetically ordered nematic phase of FeSe, a strain-enhanced orthorhombicity is expected to enhance the anisotropy of the spin fluctuations \cite{Ghini2021,He2018} and their effect on transport anisotropy \cite{Onari2017,Fernandes2013}, reflecting the propensity of the system to undergo a putative stripe-type magnetic transition. This then can explain both the decreasing magnitude of the elastoresistivity below $T_s$ as well as the distinct proportionality in the detwinning and post-detwinning strain regimes (see Supp. Fig. S11). A strain-transport study of FeSe using a more direct probe of spin fluctuations with a tunable strain state would be needed to confirm this scenario.   \\

\textbf{Discussion}\\

Due to the intertwined nature of spin and orbital degrees of freedom, two general routes have been invoked to explain the nematicity in FeSe. In the spin-nematic picture, divergent orbital-selective spin fluctuations drive the nematic ordering and consequently induce a splitting between $|3d_{xz}\rangle$ and $|3d_{yz}\rangle$ bands. Our observation of (i) a large nematic-phase spontaneous orbital polarization, (ii) its divergence behavior above $T_s$ and (iii) the direct correspondence between the orbital polarization and the transport anisotropy above $T_s$ overall suggests that nematicity is instead driven by orbital order.

In this case, the increasing Fermi surface anisotropy below $T_s$ enhances strongly anisotropic spin fluctuations (as evidenced by our elastoresistivity analysis) which can themselves act as a mechanism for further momentum-dependent evolution of the Fermi surface \cite{Pfau2019, Fanfarillo2016, Rhodes2020}. The change of the Fermi surface topology and orbital content at the hole and electron pockets observed by ARPES \cite{Yi2019,Pfau2019}, the associated suppression of $B_{1g}$ charge fluctuations in Raman \cite{Massat2016,Udina2020}, the increase of the spin-relaxation rate in NMR \cite{Baek2015}, the abrupt sign-change in the Hall coefficient \cite{Ghini2021, Sun2017}, and the unusual sign-changing elastoresistivity \cite{Watson2015,Hosoi2016, Ghini2021}, all manifesting only {\it below} $T_s$, may be the clearest signs of this effect. This sequence of events, where orbital-dependent spin fluctuations are triggered below $T_s$ by the onset of nematicity driven by spontaneous orbital order may reconcile the apparently disparate conclusions reached by previous studies and is consistent with our experimental observations. Indeed, inelastic neutron scattering has shown that the spin frustration between N\'{e}el order (with $C_4$ rotational symmetry) and stripe-type order (with $C_2$ rotational symmetry) is partially lifted by the nematic transition and its accompanying orthorhombic structural distortion \cite{Wang2016a}. Thus, our proposed notion of orthorhombicity-stabilized spin fluctuations is consistent with the available experimental evidence from more direct probes of the spin degrees of freedom \cite{Wang2016, Wang2016a, Chen2019, Lu2022}, although additional data utilizing the tunable strain methodology introduced here is required to assess how spin fluctuations modulate the properties of the orbital nematic state.

Finally, several observations stand as key indicators that nematicity and superconductivity are driven by distinct degrees of freedom in FeSe. Below the onset of superconductivity, spin fluctuations are diminished for both Co-doped BaFe$_2$As$_2$ \cite{Pratt2009} and FeSe \cite{Baek2015, Wang2016}, but the orthorhombicity is only suppressed for the former \cite{Nandi2010} while being apparently unaffected in the latter \cite{Bohmer2013}. Further, optimal superconductivity is not found in the vicinity of the nematic quantum critical point in FeSe under hydrostatic pressure \cite{Sun2016_pressure} or S-doping \cite{Hosoi2016,Baek2020}, in contrast to the iron pnictides. This may be naturally understood in our interpretation of orbital nematicity, as this regime of the phase diagram does not then correspond to the critical spin fluctuations thought to drive the unconventional pairing in the pnictides \cite{Si2016}. Instead, the presence of an orbital nematic state may affect other aspects of superconductivity in FeSe, including the anisotropy of the superconducting gap and the orbital dependence of the pairing state \cite{Hanaguri2018, Liu2018}. These results disentangle spin fluctuations from nematicity in FeSe, and thus allow a refocusing towards the more relevant part of phase diagram for optimizing chalcogenide superconductivity.

\clearpage
\section*{Acknowledgments}

We thank R. Fernandes, M. Norman, J. Analytis, M. Le Tacon, J. Pelliciari, E. Ergeçen, Q. Jiang, P. Malinowski, M. Bachmann, M. Yi and H. Pfau for discussions. We thank L. G. Pimenta Martins and Y. Lee for assistance with preliminary X-ray experiments. Work at MIT (C.A.O., J.J.S., Q.S. and R.C.) is supported by the Department of Energy, Office of Science, Office of Basic Energy Sciences, under Award Number DE-SC0019126 (resonant X-ray spectroscopy and diffraction measurements, and sample growth), by the Air Force Office of Scientific Research Young Investigator Program under grant FA9550-19-1-0063 (strain set-up development and transport measurements), and by the National Science Foundation under grant no. 1751739 (optical dichroism). The work performed at the Advanced Photon Source was supported by the US Department of Energy, Office of Science, Office of Basic Energy Sciences under contract no. DE-AC02-06CH11357. Sample preparation was performed in part at HPCAT (Sector 16) at the Advanced Photon Source, Argonne National Laboratory. J.J.S. acknowledges the support of the National Science Foundation MPS-Ascend Postdoctoral Research Fellowship under award no. 2138167. Any opinions, findings and conclusions or recommendations expressed in this material are those of the author(s) and do not necessarily reflect the views of the National Science Foundation.

\section*{Author Contributions}

C.A.O., J.J.S., G.F., Y.C., J.W.K., and P.R. conceived the project and performed all x-ray measurements. Q.S. grew the samples. C.A.O and J.J.S. prepared the samples for strain measurements, performed the optical birefringence measurements, analyzed the data and wrote the manuscript. All authors contributed to discussion of the results and commented on the manuscript. R.C. supervised the project.

\section*{Competing Interest Statement}
The authors declare no competing interests.

\clearpage

%%%%%%%%%%%%%%%%%%%%%%%%%%%%%%%%%%%%%%%%%%%%%%%%%%%%
%%%%%%%%%%%%%%%%%%%%%%%%%%%%%%%%%%%%%%%%%%%%%%%%%%%%
%%%%%%%%%%%%%%%%%%%%%%%%%%%%%%%%%%%%%%%%%%%%%%%%%%%%
%%%%%%%%%%%%%%%%%%%%%%%%%%%%%%%%%%%%%%%%%%%%%%%%%%%%
%%%%%%%%%%%%%%%%%%%%%%%%%%%%%%%%%%%%%%%%%%%%%%%%%%%%
%%%%%%%%%%%%%%%%%%%%%%%%%%%%%%%%%%%%%%%%%%%%%%%%%%%%
%%%%%%%%%%%%%%%%%%%%%%%%%%%%%%%%%%%%%%%%%%%%%%%%%%%%
%%%%%%%%%%%%%%%%%%%%%%%%%%%%%%%%%%%%%%%%%%%%%%%%%%%%
%\section*{Figure Captions}
%%%%%%%%%%%%%%%%%%%%%%
%Figure 1

\clearpage

\begin{figure}[h]
\centering
\includegraphics[width = 0.5 \columnwidth]{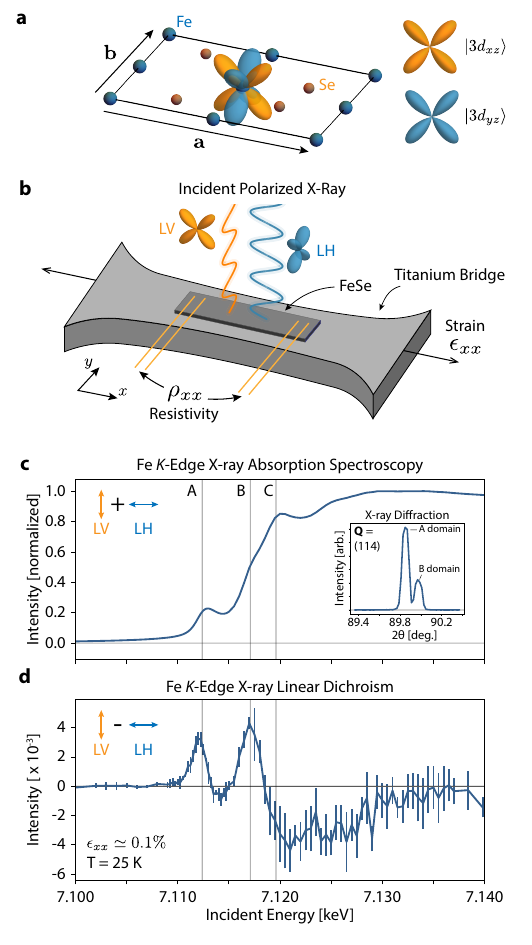}
\end{figure}

\noindent {\bf Fig. 1 | Strain apparatus and XLD spectroscopy in FeSe.} {\bf a}, The orthorhombic unit cell of FeSe, defining the $\mathbf{a}$/$\mathbf{b}$ orthorhombic axes and the orientation of the $|3d_{xz}\rangle$/$|3d_{yz}\rangle$ orbitals (A domain). {\bf b}, Schematic of the strain apparatus and sample, with polarization states (LH/LV) for normal incidence dichroism measurements. {\bf c}, Sample 1, the Fe $K$-edge XAS profile at $T = 25$ K under tension  ($\epsilon_{xx} \simeq 0.1 \%$). Inset: XRD of (114) reflection shows partial detwinning to the A domain. {\bf d}, The corresponding XLD spectrum with error bars determined from the standard deviation of repeated scans ($n = 5$, see Methods).\\

%%%%%%%%%%%%%%%%%%%%
%Figure 2
\clearpage
\begin{figure}
\centering
\includegraphics[width = \columnwidth]{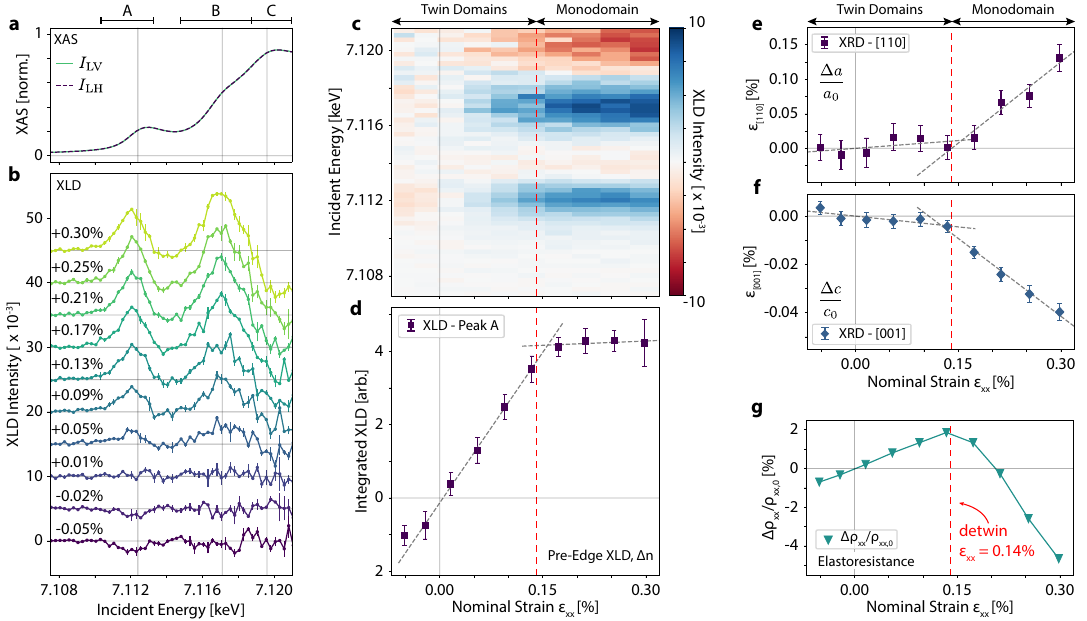}
\end{figure}

\noindent {\bf Fig. 2 | Strain-dependent X-ray measurements in the nematic phase.} X-ray measurements versus strain on Sample 1 at $T = 50$ K. {\bf a}, The Fe $K$-edge XAS profile for polarizations LH $\parallel \mathbf{a}$ and LV $\parallel \mathbf{b}$, with the integration regions for the features A/B/C denoted. {\bf b}, In-plane XLD spectra as a function of increasing strain from $-0.05\%$ (bottom) to $0.30\%$ (top). Errors bars represent standard deviation of repeated scans ($n = 2$). {\bf c}, Color map of the strain-dependent in-plane XLD spectra in {\bf b}. {\bf d}, The integrated XLD intensity of the pre-edge A ($\Delta n$) as a function of applied strain with error bars corresponding to propagated error from the XLD spectra in {\bf b}. XRD measurements of the {\bf e}, $[110]$ (parallel to applied strain) and {\bf f}, the $[001]$ (out-of-plane) lattice strains versus nominal linear strain $\epsilon_{xx}$. Error bars represent the standard error from Gaussian fits to XRD peaks. {\bf g}, Elastoresistance measurements recorded simultaneously with XLD and XRD data, showing a sign-reversal at the detwinning point, $\epsilon_{xx} \simeq 0.14\%$, denoted by vertical red lines in panels {\bf c}-{\bf g}. All measurements reported in {\bf a}-{\bf g} are taken at identical strain conditions.\\

%%%%%%%%%%%%%%%%%%%%%%%%%%%%%%%%%%%%%%
%Figure 3
\clearpage
\begin{figure}
\centering
\includegraphics[width = 1 \columnwidth]{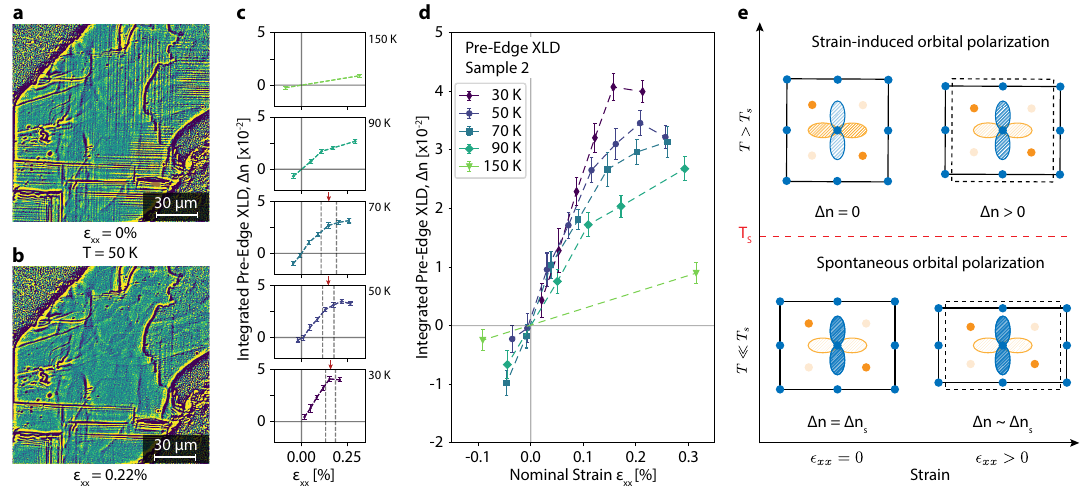}
\end{figure}

\noindent {\bf Fig. 3 | Spontaneous orbital polarization across the nematic transition.} Optical birefringence images of the nematic domains at $T = 50$ K on Sample 2 at {\bf a}, $\epsilon_{xx} = 0\%$ and {\bf b}, $\epsilon_{xx} = 0.22\%$ strain showing the full detwinning. Scale bars are $30$ $\mu$m. {\bf c}, Temperature and strain-dependent pre-edge XLD ($\Delta n$) across the nematic transition in Sample 2. For $T < T_s$, the approximate detwinning points are indicated by red arrows, determined from independent XRD and optical birefringence measurements (see Methods and Supp. Figs. S11, S12). Vertical dashed grey lines indicate the range of uncertainty in the detwinning point. {\bf d}, The same data in {\bf c} plotted together to highlight the temperature- and strain-dependence. XLD error bars are as defined in Figures 1 and 2. {\bf e}, Above the nematic transition ($T>T_{s}$, top), the orbital polarization is only nonzero with applied strain. Far below the nematic transition ($T \ll T_{s}$, bottom), the orbital polarization is spontaneously nonzero, fully saturated, and is unchanged with applied strain beyond full detwinning, indicating it is not driven by the lattice distortion. \\

%%%%%%%%%%%
%Figure 4
\clearpage

\begin{figure}
\centering
\includegraphics[width = 0.5 \columnwidth]{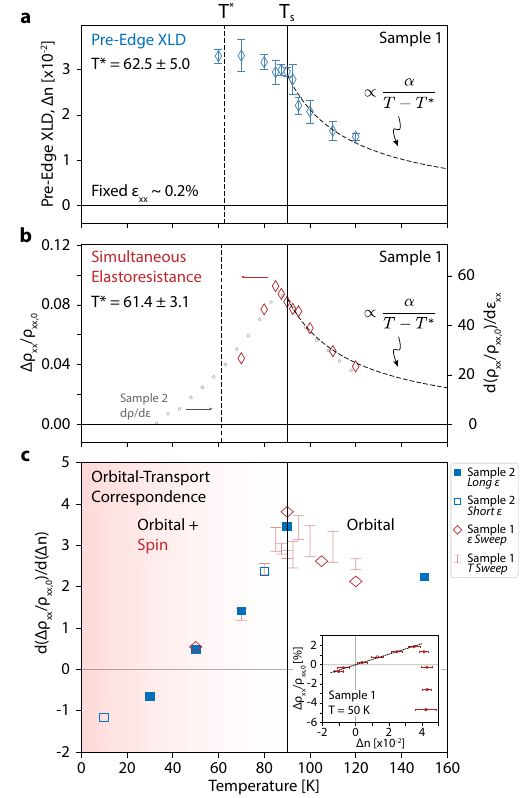}
\end{figure}

\noindent {\bf Fig. 4 | Divergent orbital polarizability and orbital-transport correspondence.} {\bf a}, $\Delta n$ and {\bf b}, the simultaneous elastoresistance (red diamonds, left axis) versus temperature across $T_s$ at a fixed $\epsilon_{xx} \simeq 0.2 \%$ for Sample 1. Dashed curves are Curie Weiss fits to the data at $T \geq T_s = 90$ K, indicating CW temperatures of $T^* \simeq 62.5 \pm 5$ K and $61.4 \pm 3$ K, respectively. Also shown in {\bf b} is the longitudinal elastoresistivity coefficient about zero strain in Sample 2 (grey circles, right axis - see Supp. Fig. S11) for comparison. {\bf c}, Ratio of the normalized resistivity to the pre-edge XLD versus temperature, taken from the slope of the linear fit of normalized resistivity at $T = 50$ K from {\bf c} (inset), additional temperatures in Sample 1 (Supp. Fig. S17, red diamonds), equivalent data from Sample 2 (from Fig.3{\bf c} and Supp. Figs. S16/S18, blue squares), and from the fixed-strain temperature sweep in Sample 1 (Fig.4{\bf a,b} (red bars). Inset: Normalized resistivity versus XLD in Sample 1 at $T = 50$ K using data from Fig. 2{\bf d}/{\bf g}. Error bars in {\bf a}/{\bf c} are determined from the statistical errors of the XLD signal, and are the size of the data point where not indicated.\\

\clearpage

%\bibliographystyle{naturemag}
%\bibliography{FeSe_XLD_references_corrected_20230101}

\clearpage

\section*{Methods}

\noindent \textbf{Sample preparation.} Single crystals of FeSe with typical dimensions of 2.0 x 2.0 x 0.05 mm were grown using the KCl-ACl$_3$ flux method as described in Ref. \cite{Wang2016}. The resulting samples were cut with a $\lambda = 1064$ nm laser cutter (spot size of 10 $\mu$m) into rectangular bars with dimensions of 1.0 x 0.15 mm with the long axis along the tetragonal [110] (orthorhombic {\bf a}) direction. The samples were then mounted onto pre-fabricated Titanium platforms (following the specifications outlined in Refs. \cite{Bartlett2021} and \cite{Park2020}) using Stycast 2850LT epoxy, cured at 80$^\circ$ C for 12 hours. After the epoxy was cured, the sample surface was cleaved several times to reduce the thickness to $\simeq$ 25 $\mu$m  and 25 $\mu$m gold wires were affixed near the ends of the long axis of the sample [main text Figure 1{\bf b} and Figure S1] in a four-wire geometry using Epotex H20E silver epoxy. The silver epoxy was cured in a nitrogen-filled glovebox at 120$^\circ$ C for 15 minutes for Sample 2 and at 80$^\circ$ C in ambient conditions for 2 hours for Sample 1. The Ti platform with the wired FeSe bars was then fastened to a Razorbill CS130 strain cell. \\

\noindent \textbf{XAS, XLD and XRD Measurements.} X-Ray absorption and diffraction measurements were performed at the 4-ID-D endstation of the Advanced Photon Source, Argonne National Laboratory. The strain cell was mounted into an Advanced Research Systems Displex closed-cycled cryostat with a base temperature $T \simeq 10$ K. The cryostat was mounted in a 6-circle diffractometer to allow sample manipulation for XRD, as well as precise alignment of the incident beam to the surface normal of the sample for in-plane XLD measurements. XAS spectra were recorded in partial fluorescence yield at the Fe-K$\alpha$ emission line in a quasi-backscattering geometry with a Hitachi Vortex detector. The incident energy was varied across the Fe $K$-edge resonance, and at each incident energy the polarization was rapidly switched between linear polarization states using diamond phase plates in the sequence LV/LH/LH/LV in order to determine the linear dichroism spectrum with high signal-to-noise ratio.

All reported XAS spectra were normalized to the main edge jump using reference XAS/XLD spectra acquired up to 7.140 keV. The reported XLD is calculated as the direct difference of the normalized XAS spectra at different polarizations. For the high-statistics near-edge XLD spectra used for Figure 2 and Figure 3 in the main text, 2-3 spectra were acquired back to back and averaged, in the energy range of 7.107 - 7.121 keV with a step size of 250 meV. The standard deviation of these subsequent scans is used for the error bars of the XLD data. XRD measurements were performed with an incidence energy above the Fe $K$-edge resonance, with an energy of 7.200 keV and 7.400 keV for Sample 1 and Sample 2, respectively. The X-ray measurements are performed with a beam size of $50 \times 50$ $\mu$m$^2$ and are bulk sensitive, and thus average over multiple twin domains when present \cite{Tanatar2016} (Fig. 3{\bf a}/{\bf b}). \\

\noindent \textbf{Strain-Temperature Sweep Procedures.} All data reported were taken by sweeping the voltage of the piezo stacks of the strain device continuously from maximum compression to maximum tension. Before changing temperature, the voltage on the piezo stacks was fixed to zero. This ensured a consistent nominal zero-strain capacitance reading throughout the experiment. Once at target temperature, the sample would be initialized by moving to maximum tension and back to maximum compression before initializing the sweep. This consistent initialization procedure accounts for any effects of detwinning hysteresis or hysteretic effects of the piezo-actuators/Ti bridge, ensuring consistent and comparable strain-dependent measurements throughout the experiment. We further note that as we directly control the total length of the sample via applying stress to the titanium platform, we do not directly observe hysteresis in the detwinning process, in contrast to the methodology involving a sample suspended over a gap as in Ref. \cite{Sanchez2021}.

For the strain sweep data in Figure 2, the sample was initialized with the above procedure at $T = 50$ K and strain was increased monotonically from maximum compression to tension. At each strain point, XAS/XLD spectra at normal incidence and XRD measurements of the $(114)$ and $(004)$ were performed at an identical strain condition before moving to the next increased tensile strain point. The resistivity values were stable at fixed strains and the reported resistance values were averaged over the duration of corresponding XLD measurements.

For the XLD temperature dependence recorded in Sample 2 (Fig. 3), the measurements were completed from the same, single cooldown of the cryostat, starting at low temperature (10 K) and increasing monotonically to the highest measured temperature (150 K). Reported resistance measurements were again averaged over the duration of the XLD measurements, and thus the reported quantities are under identical strain conditions. XRD measurements were also performed in Sample 2. These were performed on subsequent strain loops at the same temperature conditions, immediately after the XLD/resistance strain loops. The sample was consistently initialized, and separate strain sweeps following the same compressive-to-tensile strain loop as the XLD measurements were repeated for separate measurements of the $(114)$ and $(004)$ reflections.\\

\noindent \textbf{Determination of Nominal Strain.} All reported data are plotted versus nominal linear strain $\epsilon_{xx}$. For XLD measurements, the nominal zero-strain was determined by the capacitance readout of the strain cell gap corresponding to the interpolated value for zero XLD signal, signaling a fully twinned sample below $T_S$, or an unstrained tetragonal phase above (see Supp. Fig. S15). Since XLD is bulk sensitive, this provides the most consistent measure of the true zero strain value. The conversion of the change in the strain cell gap distance with respect to this zero, monitored by change in the capacitance across the gap, was determined by strain transmission data at $T = 90$ K of the [110] lattice parameter, calculated from the combined XRD of the (114) and (004) reflections. The temperature $T = 90$ K was chosen for strain transmission data since the lack of orthorhombic domains results in a linear strain transmission, as opposed to the highly non-linear effects observed within the nematic phase (Main text, Fig. 2{\bf e}/{\bf f}) due to details of the detwinning process.\\

\noindent \textbf{Optical Birefringence Measurements.} Optical birefringence and additional elastoresistivity data were acquired on Sample 2 with identical sample preparation conditions (no change in the strain cell, mounting or wiring). Polarized images were acquired using a mono-chrome camera with a broadband LED light source which was passed through a $\lambda = 600$ nm long-pass filter (ThorLabs) in order to increase the birefringence contrast \cite{Tanatar2016}. The incident light was polarized with a broadband Glan-Thompson polarizer (ThorLabs) along the tetragonal [100] direction (bisection the orthorhombic axes). To resolve the birefringence-induced polarization rotation, another Glan-Thompson polarizer was placed before the camera and detuned by $\simeq 1^\circ$ from the cross-polarized configuration (with respect to the polarizer) in order to optimize the birefringent domain contrast. The images were recorded using both 20x and 50x objectives (Olympus).

\end{document}

\title{Spontaneous orbital polarization in the nematic phase of FeSe -- \textit{Supplementary Information}}

\affiliation{Department of Physics, Massachusetts Institute of Technology, Cambridge, MA 02139, USA}
\affiliation{Advanced Photon Source, Argonne National Laboratory, Lemont, IL 60439, USA}
\affiliation{These authors contributed equally to this work}

\author{Connor A. Occhialini}
%\email{caocchia@mit.edu}
\affiliation{Department of Physics, Massachusetts Institute of Technology, Cambridge, MA 02139, USA}
\email{These two authors contributed equally}
\affiliation{These authors contributed equally to this work}

\author{Joshua J. Sanchez}
%\email{sanchezx@mit.edu}
\affiliation{Department of Physics, Massachusetts Institute of Technology, Cambridge, MA 02139, USA}
\affiliation{These authors contributed equally to this work}

\author{Qian Song}
\affiliation{Department of Physics, Massachusetts Institute of Technology, Cambridge, MA 02139, USA}

\author{Gilberto Fabbris}
\affiliation{Advanced Photon Source, Argonne National Laboratory, Lemont, IL 60439, USA}

\author{Yongseong Choi}
\affiliation{Advanced Photon Source, Argonne National Laboratory, Lemont, IL 60439, USA}

\author{Jong-Woo Kim}
\affiliation{Advanced Photon Source, Argonne National Laboratory, Lemont, IL 60439, USA}

\author{Philip J. Ryan}
\affiliation{Advanced Photon Source, Argonne National Laboratory, Lemont, IL 60439, USA}

\author{Riccardo Comin}
\email{rcomin@mit.edu}
\affiliation{Department of Physics, Massachusetts Institute of Technology, Cambridge, MA 02139, USA}

\date{\today}

\maketitle

\tableofcontents

\clearpage

\section*{Supplemental Text}

\subsection*{Characterization of XLD Signal} 
Here we provide additional XLD data to corroborate our measurement of XLD purely within the Fe-Fe rectangular lattice plane. FeSe is tetragonal above the nematic phase, and thus at all temperatures possesses a large structural anisotropy between the out-of-plane direction ($\mathbf{c}$ axis) and the $\mathbf{a}$-$\mathbf{b}$ plane. We report measurements assessing this effect in Figure S2, acquired on Sample 1 at $T = 25$ K. In order to measure the out-of-plane component, we rotated the sample about the $\mathbf{a}$ axis so that the incoming beam makes an angle of $\theta = 50^\circ$ with respect to the surface normal of the sample. In this geometry, LV polarization acquires a finite projection along the $\mathbf{c}$ axis while LH polarization remains oriented along $\mathbf{a}$. The resulting XLD spectrum, which we refer to as the ``out-of-plane'' XLD, is shown in Figure S2{\bf b}, showing a large dichroism of the order $10^{-1}$ which is visually inferred from the direct XAS profiles in LH/LV polarizations in Fig. S2{\bf a}. For comparison, we also show the normal incidence XLD spectrum in a partially detwinned sample with LV/LH parallel to $\mathbf{b}$/$\mathbf{a}$ orthorhombic (Fe-Fe direction) axes, respectively, which we refer to as the ``in-plane'' XLD spectrum. 

While the in-plane XLD is over an order of magnitude weaker, the spectral shape is distinct from the out-of-plane XLD corroborating a distinct origin, ruling out any small misorientation of the sample and parasitic contributions from the out-of-plane spectrum into the in-plane XLD spectra reported here. In the case of our measurement, the surface normal was aligned to the incoming wavevector to an accuracy exceeding $1^\circ$, using a diffractometer and the experimentally determined orientation matrix of the sample. XRD was also used to accurately align the orthorhombic axes to lie parallel to LV/LH polarizations for the in-plane XLD measurements. 

We also acquired in-plane XLD spectra in the fully detwinned state of Sample 1 at $T = 50$ K in two sample orientations to associate the measured XLD spectra to an intrinsic response of the sample. We first acquired XLD spectra at an angle $\phi = 0^\circ$ corresponding to LV $\parallel \mathbf{b}$ and LH $\parallel \mathbf{a}$, and then rotated the sample to $\phi = 90^\circ$ in a plane perpendicular to the incoming beam, with the second state corresponding to LV $\parallel \mathbf{a}$ and LH $\parallel \mathbf{b}$. The measurement procedures and the strain state of the sample were held under identical conditions. The two spectra are shown in Figure S3. The two spectra are reversed in sign with respect to each other and quantitatively similar in magnitude and spectral shape. The results of Fig. S2 and S3 confirm that the in-plane XLD is genuine and is not resulting from an extrinsic effect related to misorientation or normalization errors in the incoming polarization states. For the strain-dependent curves in main text Fig. 2{\bf b}, this also confirms there is not any constant offset from the instrumentation in a fixed geometry as was done for all data reported in the main text. For the XLD in main text Fig. 2{\bf a}-{\bf d}, we also show the strain-dependence of the XLD spectra resolved into the individual spectral features A/B/C (as identified in Figure S2) in Figure S4.\\

\subsection*{FDMNES Calculations} 

FDMNES calculations were performed using the experimentally determined lattice parameters at base temperature ($ a = 5.331$ {\AA}, $b = 5.305$ {\AA}, $c = 5.485$ {\AA}) using the $Cmma$ space group for orthorhombic FeSe \cite{Bunau2009}. The Se atomic height of $z = 0.2673$ was used. The Fermi level was set self-consistently and the convergence with respect to cluster radius was checked up to 12 {\AA} within the Muffin Tin approximation. Only small changes were found above 7 {\AA} cluster radius in both the simulated XLD and XAS. For the reported calculations, a cluster radius of $7.3$ {\AA} was chosen in order to include the full shell of 11${}^{th}$ nearest neighbor Fe atoms. This value is consistent with previous reports for FDMNES calculations in Fe-based superconductors \cite{Lafuerza2017}. Unless explicitly specified, both quadrupole and dipole transition intensities are included in the calculations. 

We compare the polarized XAS and XLD spectra output from FDMNES to the experimental data in Figure S5. For comparison to the experimental data, the Fermi level was set to an edge energy of $E = 7.1124$ keV and the energy axes of the calculations were stretched by $15\%$, as argued in Ref. \cite{Lebert2018}. A satisfactory agreement is found for the XAS spectra, as well as the overall spectral shape and magnitude of both the in-plane and out-of-plane XLD spectra. The experimental data and the calculated curves are both normalized to the main edge jump using the same procedure. The in-plane experimental data in Fig. S5{\bf b} corresponds to Sample 1 at $T = 50$ K and is in the fully detwinned state, while the out-of-plane data in Fig. S5{\bf d} is from Sample 1 at $T = 25$ K. The out-of-plane XLD spectrum is purely structural in origin and is well-captured by the calculations, resulting from the large anisotropy determined from the high-temperature tetragonal structure. The XLD magnitude is slightly overestimated by the calculation. The expected structural contributions at higher energy features (peak B/C) for the in-plane XLD are also well reproduced, although again overestimated compared to experiment. For the pre-edge peak A, the spectral shape is not captured well, and the overall intensity is far underestimated. We note that without the introduction of electronic correlations, the disagreement could be due to an improper treatment of the $3d$ orbitals by a single-particle code such as FDMNES \cite{Lebert2018}; however, general agreement has been found between multiple scattering codes and the more strongly correlated $L_{2,3}$ XAS spectra in Fe-based superconductors \cite{Yang2009,Chen2010,Kim2013}. This suggests a non-structural mechanism that is not captured by the FDMNES calculations to explain the pre-edge XLD spectrum, which is attributed to properties of the localized orbitals derived from Fe $3d$, Se $4p$ and an admixed Fe $4p$ character, as has been recently suggested \cite{Fguieredo2022}.

To support this further, we investigate the origin of the pre-edge peak in more detail. Figure S6 shows the same calculation of the in-plane XLD, but resolved into the dipolar transitions ($1s \to 4p$) and the quadrupolar transitions ($1s \to 3d$). The quadrupolar matrix elements are calculated for wavevector $\mathbf{k_i} \parallel \mathbf{c}$ as in the normal incidence measurements for the in-plane XLD configuration executed experimentally. The quadrupole transition intensity is largest at the pre-edge, which may be anticipated as this corresponds to the region of the unoccupied Fe $3d$ states just above the Fermi level. However, even at the pre-edge, the direct quadrupolar transition is approximately a factor of 40 weaker than the dipole contribution in both the XAS and XLD spectra. Thus, the signal measured in experiment is almost purely dipole in origin as has been concluded previously for the Fe $K$ pre-edge \cite{Lebert2018, Chen2012, Lafuerza2017, Pelliciari2021, Simonelli2012, Joseph2010}. The increase in the dipolar intensity is due to the both local tetrahedral environment and hybridization with the ligand Se $4p$ states. In the tetrahedral crystal field, a local on-site mixing of the Fe $4p$ orbitals with the Fe $3d$ states becomes allowed (discussed more below). This corresponds to an orbital localization effect as suggested in Ref. \cite{Fguieredo2022} and is supported by both experiments \cite{Joseph2010, Chen2012} and Wannier projection of the $3d$ orbitals \cite{Lee2009}.

As a final check, we performed calculations as a function of the structural orthorhombicity $\delta$. In experiment, the orthorhombicity at base temperature is approximately $\delta = 0.24\%$. The in-plane XLD spectra for several $\delta$ are shown in Figure S7. As can be seen, the predicted structural XLD is linear in $\delta$ at each XAS spectral feature. This is in contrast with the behavior of the pre-edge found in experiment, both from the strain dependence (main text Fig. 2/3) and the temperature dependence (main text Fig. 3/4). As a function of strain below $T_s$, only the higher energy features continue to increase beyond the detwinning point as the structural anisotropy is increased beyond the spontaneously ordered value. Furthermore, when lattice orthorhombicity is externally imposed by strain above $T_s$, the relative ratio of the high-energy (B/C) and the pre-edge (A) features changes with respect to the nematic phase, with similar magnitudes in the higher energy features but a suppressed intensity at the pre-edge (see main text Figure 4{\bf a} and Fig. S14 below). These aspects imply a spontaneously ordered orbital polarization that is independent from the lattice that arises below $T_s$.

The overall interpretation of the pre-edge sensitivity to the $3d$ states is shown in Figure S8, provided (a) the dipolar origin of the in-plane XLD signal and (b) the localization effects beyond single-particle approximations suggested in Ref. \cite{Fguieredo2022}. The local tetrahedral environment around the central Fe atom breaks the local inversion symmetry which allows an on-site hybridization of Fe $3d$ and Fe $4p$ states. This effect has been experimentally verified \cite{Chen2012} and is the reason why the Fe $K$ pre-edge of tetrahedrally coordinated Fe is more intense, due to the increase of the dipolar $1s \to 4p$ contribution which is forbidden in higher symmetry (e.g. octahedral) compounds \cite{Yamamoto2008}. Even with small mixing, the dipole transition amplitude is several orders of magnitude larger than the quadrupole one, and therefore quickly becomes dominant. It has been argued recently that this corresponds to an effective orbital localization of the Fe $3d$ states by reducing the Fe $3d$ anti-bonding character with respect to hybridization with the neighboring Se $4p$ states \cite{Fguieredo2022}. This effect has been shown to be common in the Fe-based superconductors, but is particularly pronounced in the chalcogenides as inferred through comparison of the Se $K$- and the Fe $K$-edge XAS polarization dependence \cite{Fguieredo2022, Joseph2010}.

In the case of iron pnictides and chalcogenides, Wannier orbital analysis of the $3d$ states near the Fermi level confirms that the Fe states in the $t_{2g}$ sector hybridize most strongly with orbitals of $p$ character that are oriented perpendicular to the plane of the given $3d$ state \cite{Lee2010}. This is shown for the case of the $|3d_{yz}\rangle$ orbital in Fig. S8{\bf a}, which mixes most strongly with the $|4p_x\rangle$ orbital leading to a hybrid orbital at the far left where the $3d$ lobes are deviated towards the nearest neighbor Se atoms (see Fig. S8{\bf b}). In this situation, the dipolar transition at the pre-edge corresponds to a final state of the $|4p_{x}\rangle$ component of the hybridized orbital, which may be selected with linear polarized light along the $\mathbf{a}$ direction (Figure S8{\bf b}, right). Also shown is the analogous situation for the $|3d_{xz}\rangle$ orbital, which mixes strongest with the $|4p_{y}\rangle$ state and may be accessed through dipolar transitions with linear polarization along $\mathbf{b}$ (Figure S8, left). With these assignments, the positive XLD at the pre-edge (corresponding to $I_{LV} - I_{LH}$ with LV/LH parallel to $\mathbf{b}/\mathbf{a}$, respectively) corresponds to a higher occupation of the $|3d_{yz}\rangle$ orbital, as argued in the main text. While the mixing of $4p$ character may be sensitive to the bond-lengths, the saturating XLD signature with strain past the detwinning point suggests that the change in hybridization is a subdominant effect compared to the spontaneous orbital occupation mismatch of the $3d$ orbitals. \\

\subsection*{Additional XRD Data in Sample 1}

Fig. S9{\bf a}/{\bf b} shows the raw x-ray diffraction scans as a function of strain for the (114)/(004) Bragg reflections, respectively, which were used to extract the strain-dependent lattice parameters in main text Figure 2. All XRD data were acquired with an energy of $7.2$ keV in Sample 1. The peak positions were determined using Gaussian fits, shown as dashed black lines. We recall the procedure for the strain loop on Sample 1. All data was collected on the same strain loop; first, the strain state was initialized, followed by XRD of the (004), then the (114) peaks, and finally the XLD spectra were recorded. Only after all data collection was the strain state increased monotonically in the tensile direction.

We note that in the XRD data in Fig. S9 that the twin peak corresponding to the compressive domain (with shorter {\bf b} axis aligned to the strain direction) disappears completely near the detwinning point of $\epsilon_{xx} \simeq 0.14\%$.  The relative domain population (judged by the relative peak intensities) versus strain appears to be non-linear. However, provided the linearity of the elastoresistance and the XLD in the detwinning regime, as well as supporting evidence from birefringence experiments discussed below in Sample 2, the evidence is strong for linear domain population evolution versus strain, integrated throughout the bulk of the sample. The apparent discrepancy between these measurements and the XRD data likely results from our use of a point detector which selects a small region of the total diffraction intensity due to the finite crystal mosaicity common for weakly bonded and layered samples. Thus, while the $2\theta$ positions of the peaks are robust, the absolute and relative intensities of the diffraction peaks are not necessarily representative of the bulk of the sample, but rather of a localized region which is selected by our choice of sample angle. These aspects only effect the XRD measurement which is highly sensitive to orientation, while the bulk probes of XLD and elastoresistance are not sensitive to small grain misorientations. Despite these considerations, we note that the apparent detwinning point from disappearance of the compressive domain peak is consistent with the multiple independent signatures, including the slope changes in the lattice parameters, elastoresistance and XLD data as shown in main text Figure 2. These considerations are also important to keep in mind below when discussing analogous XRD data recorded in Sample 2. 

XRD was also recorded at $T = 120$ K on Sample 1 as a function of strain, shown in Figure S10. The twin domains are clearly absent and both the $[110]$ and the $[001]$ lattice parameters (shown in Fig. S10{\bf c}/{\bf d}, respectively) evolve linearly with strain, in clear qualitative distinction compared to the lattice behavior at $T = 50$ K (main text Figure 2). The simultaneous integrated pre-edge XLD and elastoresistance (Fig. S17{\bf c}/{\bf f}, respectively) are also linear over this region, as discussed below.\\

\subsection*{Optical Domain Imaging and Additional Elastoresistance Data in Sample 2} 

In addition to the in-situ strain, transport, XLD and XRD experiment performed on Sample 2 presented in the main text Figure 3, we executed additional experiments on the same sample using in-situ strain in conjunction with optical birefringence to image the nematic domains \cite{Tanatar2016} while collecting simultaneous elastoresistance data.

Images were taken across a range of large tensile and small compressive strain values. Figure S11{\bf a}/{\bf b} shows polarized images of Sample 2 at 50 K under zero strain and large tensile (0.22\%) strain, respectively. Dark (bright) lines indicate the A (B) structural domains with twin boundaries along the tetragonal axes, where the domains are defined as having the (longer) {\bf a} or (shorter) {\bf b} lattice constant aligned to the stress axis, respectively, with corresponding longitudinal resistivities $\rho_{a}$ and $\rho_{b}$. Note that the sample surface has visible layer terracing which manifests as dark lines in each image which are not related to the thinner, lighter lines corresponding to the twin domains. This surface terracing does not affect the x-ray measurements which are bulk sensitive. Under large tension, the sample is detwinned towards the A domain and the corresponding twin domain lines disappear (Fig. S11{\bf b}). The domain populations were analyzed as a function of strain within a homogeneous region denoted by the white box in Fig. S11{\bf a}, which was possible for temperatures below 80 K. These estimates were then combined with the XRD results in Fig. S12 to estimate the detwinning ranges utilized for the XLD strain-dependent data in main text Figure 3 and Supplemental Figure S13.

Simultaneously recorded resistivity data are plotted in Fig. S11{\bf e} across a large temperature range above and below the nematic transition ($T_{s} = 90 K$). At all temperatures, we calculated the slope of the resistivity about the zero strain point, which diverges towards the transition from above $T_{s}$ (gray) and diminishes with cooling below $T_{s}$ (red). From the birefringence data, we determined the detwinning point at all temperatures between $40$ K and $80$ K and calculated the resistivity slope at strains beyond full detwinning (blue). Notably, this strain-induced elastoresistivity crosses zero near $60$ K, while the spontaneous resistivity anisotropy is found to cross zero near $40$ K (Main text, Figure 4). The elastoresistance data about zero strain shown in Figure 11 {\bf f} is used for the comparison data for the continuous temperature sweep in main text Fig. 4{\bf b}.\\

\subsection*{Additional Temperature- and Strain-Dependent Data in Sample 2} 

Figure S12 shows XRD data of the (114) reflection versus strain on Sample 2. All XRD data in Sample 2 were recorded with an incident energy of $E = 7.4$ keV. These data were recorded on separate (immediately preceding), but identically prepared strain loops as the XLD data presented in main text Figure 3 (in contrast to Sample 1, a separate strain loop was used for XLD and XRD data). Additional data is also shown for $T = 10$ K and $T = 80$ K for which shorter XLD loops were performed (discussed below). The twin domains are apparent with clear strain-detwinning induced by increasing tensional strain from bottom to top in each plot, while the twin domains disappear at $T = 90$ K and a linear shift of the lattice constant is visually inferred. The zero strain point was determined by the zero-crossing of pre-edge XLD. To quantify the detwinning, we plot the total intensity of each twin domain peak as a function of strain for the A domain (red) and the B domain (blue), in the bottom row of Fig. S12. We acknowledge that the apparent location of equal domain populations (which should correspond to zero applied strain) does not agree with the value inferred from the XLD for all temperatures. We attribute this to the uncertainty in the total diffraction intensity due to the detection scheme as discussed above in the context of Sample 1.

At $T = 10$ K, we have insufficient strain range to completely detwin, and instead a nearly linear change in domain populations in favor of the tensile A domain is seen over the full strain range. As temperature is increased, we see clear signs of nearly complete detwinning. We find that deep within the nematic phase, there appears to be residual pinned minority (B) domains which remain despite the increasing applied strain, and the domain intensity saturates towards a non-zero value at large tensile strain. Similar behavior was also seen in other detwinning experiments on Fe-based superconductors \cite{Sanchez2021}. As we heat towards to the transition, the presence of these pinned domains becomes less apparent and are not visible by $T = 80$ K.

Provided this behavior of the data, we estimate a range of tensile strain over which the detwinning process is nearly complete based on the saturation point (change in slope of the intensity) of the minority B domain (blue curves in the bottom row of Fig. S12). These are indicated for $30 \leq T < 90$ K as vertical dashed grey lines. The uncertainties in the detwinning point are on the order of $\pm 0.02\%$ which are set by the coarse strain step-size used. We note that the presence of more pinned domains at low temperature compared to higher temperatures acts to suppress the net XLD at low temperatures. Meanwhile an increasing trend is still observed as temperature is decreased, despite the increase of pinned domains. Thus, the increasing trend of the XLD (main text Fig. 3{\bf d} and Fig. S13{\bf f}) is robust and would only be strengthened if the pinned domains were fully accounted for. Provided the uncertainties in the true relative domain volume fractions, we do not attempt any corrections and instead present only the raw data.

The detwinning ranges from the XRD data (Fig. S12) were combined with detwinning estimations using wide-field polarized microscopy (Fig. S11) to estimate an overall detwinning range for each temperature $T < T_s$. We note that all signatures of the detwinning process give a consistent estimation of the detwinning point within uncertainties. These composite detwinning ranges (uncertainty ranges of where the detwinning point occurs) are used in Fig. S13 to discuss the trend of the pre-edge XLD and the simultaneous elastoresistance versus strain data (as shown in main text Fig. 3 and Fig. S16{\bf a}-{\bf c}).

We observe that for $T < T_s$ data in Fig. S13 that the independently determined detwinning ranges (vertical dashed grey lines) agree with the position of the observable slope change in both the XLD and the elastoresistance data, in agreement with the conclusions of main text Figure 2 for Sample 1 at $T = 50$ K. The errors in the detwinning point are used to estimate the spontaneously ordered orbital polarization and associated uncertainties in Fig. S13{\bf f}. We used the detwinning ranges to separate the data into two fitting regions, one within the twin domain region (indicated at fit range (1)) and post-detwinning (fit range (2)) which are linearly fit. This is used to estimate the strain-susceptibility of the orbital polarization post-detwinning (for $T< T_s$), shown in Fig. S13{\bf g}. This is supplemented by the estimated strain-susceptibility around zero strain for $T \geq T_s$ which is determined by the fitting range (1) indicated at $T = 90$ K and $150$ K. We acknowledge that this should be considered an estimate only, as there is insufficient data in the post-detwinning range to make an accurate determination. However, we wish to underscore the qualitative behavior of the data where the susceptibility is quenched below the transition as the spontaneous orbital polarization saturates, reminiscent of the magnetization analogy used in the main text.\\

\subsection*{Fixed-Strain Temperature-Dependent XRD and XLD Data in Sample 1} 

Here, we provide additional XRD and XLD data for the continuous temperature sweep at fixed strain in Sample 1 which is shown in main text Figure 4{\bf a}/{\bf b}. This data was acquired over two runs: the first was prepared to $\epsilon_{xx} = 0.2 \%$ strain using direct measurement of the (114) peak at $T = 120$ K (Fig. S10). The piezo voltage of the strain cell was then held constant and XLD was recorded with varying temperature on a single cooldown down to $60$ K. This data is presented in the main text Figure 4. Immediately after this, a second run was identically prepared at $120$ K and XRD was recorded for all temperatures down to $60$ K, with additional XLD at selected temperatures to check for consistency. The raw XRD and the associated lattice parameters for the second run are shown in Figure S14{\bf a}-{\bf d}. We note that upon cooling under fixed applied strain, the sample cools into a monodomain (A domain) state for all temperatures measured for $T < T_s$, as directly observable by the lack of twin domain peaks in the (114) reflection of Fig. S14{\bf a}. We note that the lattice is held in a nearly constant linear strain state within error bars ($\sim \pm 10^{-3}$ {\AA}). For reference, the unstrained $[110]$ lattice constant at $120$ K is 5.317 {\AA} and the in-plane splitting measured at $T = 50$ K is approximately $25 \times 10^{-3}$ {\AA}, showing that the variation of the strain state is not significant compared to the desired fixed-strain value of $0.2\%$ and also to the spontaneous orthorhombicity below $T_s$.

While the linear strain is determined to be nearly constant through a direct measurement of the lattice, the perpendicular reflection to determine the orthorhombicity was not directly measured due to mechanical constraints. However, paired with the fixed linear strain, we can infer a constant orthorhombicity through the structurally induced XLD responses, namely peaks B/C in comparison to the pre-edge (A). A comparison of XLD data at the highest temperature $T = 120$ K and at $T = 90$ K are shown in Fig. S14{\bf e}/{\bf f} for runs 1/2, respectively. We note a large temperature effect in the pre-edge, which is analyzed in main text Figure 4, while the higher energy features B/C are nearly constant, which is reproduced on both runs. This is consistent with a fixed lattice orthorhombicity. This also provides additional evidence that the pre-edge signature of $3d$ orbital polarization is not simply resulting from the structure, in conjunction with its distinct strain response below $T_s$ compared to the higher energy features (as discussed in Fig. S4).    \\

\subsection*{Interpretation of Strain-Dependent X-Ray and Transport Data} 

Here we explain the overall interpretation of the strain-dependent transport and XLD data. In Figure S15 we show a schematic of the strain dependent orthorhombic domain populations ({\bf a}),  net orbital polarization ({\bf b}), and longitudinal resistivity ({\bf c}). In Fig. S15{\bf d} we show the orientation of the A and B domains below $T_s$. In this figure, positive tensile strain is applied along the $x$ (vertical) direction which increasingly favors the A domain with the longer {\bf a} axis aligned to the strain direction. In this case, {\bf a}/{\bf b} always refer to the longer/shorter microscopic lattice parameters of the orthorhombic unit cells, while domain A/B refer to the respective orientation of the unit cell within a fixed lab frame defined as {\bf x}/{\bf y}.

We consider the net orbital polarization as the difference along {\bf x} and {\bf y} directions (analogous to the XLD probe with photon polarizations along {\bf x}/{\bf y}) and the longitudinal resistance measured along the strain axis $\rho_{xx}$ as executed experimentally. With these definitions, we define a positive spontaneous orbital polarization of $+\Delta n_s$ for the A domain, with the opposite sign for the B domain $-\Delta n_s$. Similarly for the resistance, we define $\rho_A$/$\rho_B$ as the resistivity $\rho_{xx}$ measured in the perfectly-detwinned A/B domain, along the {\bf a}/{\bf b} lattice parameter, respectively. The difference is related to the spontaneous resistivity anisotropy, defined as $\Delta \rho_s = \rho_A - \rho_B$.

We now consider how these quantities relate to the observed strain dependence of the measured longitudinal elastoresistance and the XLD probe of orbital polarization. At zero strain, the domain populations are equal, the net orbital polarization (XLD signal) is zero and the resistivity is $\rho_0$. As positive (negative) strain is applied the domain populations evolve linearly in favor of the A (B) domain until the detwinning strain $+\epsilon_d$ ($-\epsilon_d$) is reached. Within the ``twin domain region'' defined as $-\epsilon_d < \epsilon < \epsilon_d$, both the net orbital polarization and the longitudinal resistivity evolve linearly, interpolating between the spontaneous values of the B and A domains, with values $-\Delta n_s$/$\rho_B$ at $-\epsilon_d$ and $+\Delta n_s$/$\rho_A$ at $+\epsilon_d$. We note that these observables only hold under the assumption of a linear evolution of the domain populations and that within the twin domain region, the microscopic lattice parameters within each domain are unperturbed. These assumptions are directly confirmed in our experiment (see Figure 2). For the resistivity, the linearity would also be affected by the presence of domain wall scattering in the twin domain region, however our experiments reveal a consistent linearity in this regime across all temperatures (Figure 2, S16, S11), arguing against such a contribution.

Past the detwinning points, a monodomain state is formed and the applied strain begins affecting the microscopic lattice parameters. In this regime, a distinct dependence of the orbital polarization and resistance would be observed (defined as $d(\Delta n)/d\epsilon$ and $d\rho/d\epsilon$) corresponding to the intrinsic strain-susceptibility in the monodomain. We note that these quantities are distinct from the spontaneous values which determine the strain dependence in the twin domain regime. Thus, probing the strain tunability of the resistivity anisotropy in either the twin domain strain regime or beyond the detwinning strain measures two different quantities (i.e. $\rho_{A/B}$ vs $d\rho/d\epsilon$, effectively the transport analogue of measuring a magnetization vs a magnetic susceptibility), which here results in sharp changes in the resistivity behavior across the detwinning point as is seen in experiment. The elastoresistance post-detwinning ($d\rho/d\epsilon$) was estimated from the birefringence imaging experiments in Sample 2 in Fig. S11{\bf f} and the strain-susceptibility of the orbital polarization ($d(\Delta n)/d\epsilon)$) post-detwinning was estimated in Sample 2 in Fig. S13{\bf g}.

Most importantly, we wish to highlight the significance of the slope of the longitudinal elastoresistance to the orbital polarization within the twin domain regime. This is shown schematically in Fig. S15{\bf e} and is the subject of main text Figure 4{\bf c} and Supp. Figs. S16-S18 from the experimental data. While each quantity is linear when plotted versus strain under the assumption of a linear evolution of domain populations, when plotting the quantities against each other the relationship is expected to be linear, independent of the precise details of the detwinning process or the precise strain state of the sample. The linearity does not necessarily hold post-detwinning, which is seen most clearly in the data of Figure 4{\bf c} (inset)/Supp. Fig. S16 {\bf f} of Sample 1 at $T = 50$ K, and which signals the presence of a lattice-independent, saturating spontaneous orbital polarization in contrast to a lattice-coupled elastoresistivity. 

The slope of the resulting line within the detwinning regime provides a direct relationship between the spontaneous resistivity anisotropy and the spontaneous orbital polarization as shown in Fig. S16{\bf e} with a slope of $\Delta \rho_s/2\Delta n_s$. This is the quantity of interest which is plotted in main text Figure 4{\bf c}. We note that this slope is rigorous since both XLD and elastoresistivity are bulk probes and sensitive to similar sample volumes. Thus, even if regions of the sample are pinned or otherwise unresponsive to strain, these regions will not contribute to the net anisotropy probed by either technique and the slope will be the same, with the endpoints at the detwinning point being scaled by the responsive fraction of the sample volume. This is clearly seen in experiment as well: while the quantitative values of the bare anisotropies as measured by XLD and resistivity measurements have slight quantitative discrepancies, the slopes extracted relating these two quantities are in excellent quantitative agreement between both samples, as shown by the relative agreement in Fig. 4{\bf c}. This analysis thus gives a robust assessment of the relationship between the orbital polarization and the resistivity anisotropy and highlights the strength of a technique that can directly relate two strain-dependent quantities {\it in situ}.

Using the framework presented above, we present simultaneous measurements of the integrated pre-edge XLD intensity ($\Delta  n$) and the normalized resistivity ($\Delta \rho/\rho_{0}$) vs the nominal strain ($\epsilon_{xx}$) at fixed temperature for Sample 2 (Fig. S16 {\bf a,b}) and Sample 1 (Fig. S16 {\bf d,e}). We find that both quantities show a nonlinear dependence on nominal strain due to the combined effects of domain detwinning (revealing the spontaneous anisotropies below $T_s$) and the strain-induced orbital polarization and elastoresistivity (for strains beyond full detwinning below $T_s$, and for all strains at and above $T_s$). However, when plotting $\Delta \rho/\rho_{0}$ against $\Delta  n$ (Fig. S16 {\bf c,f}) we find that the transport is linear to the orbital polarization within the detwinning regime. The temperature dependence of this linear proportionality is shown in main text Fig. 4 {\bf c} for both samples. XLD versus elastoresistance data for additional temperatures in Sample 1 and Sample 2 are shown in Figs. S17 and S18 below.

\clearpage
%Supplemental Figures 
\section*{Supplemental Figures}
%XLD Characterization (S1-S4)

\begin{figure}[h]
\centering
\includegraphics[width = \columnwidth]{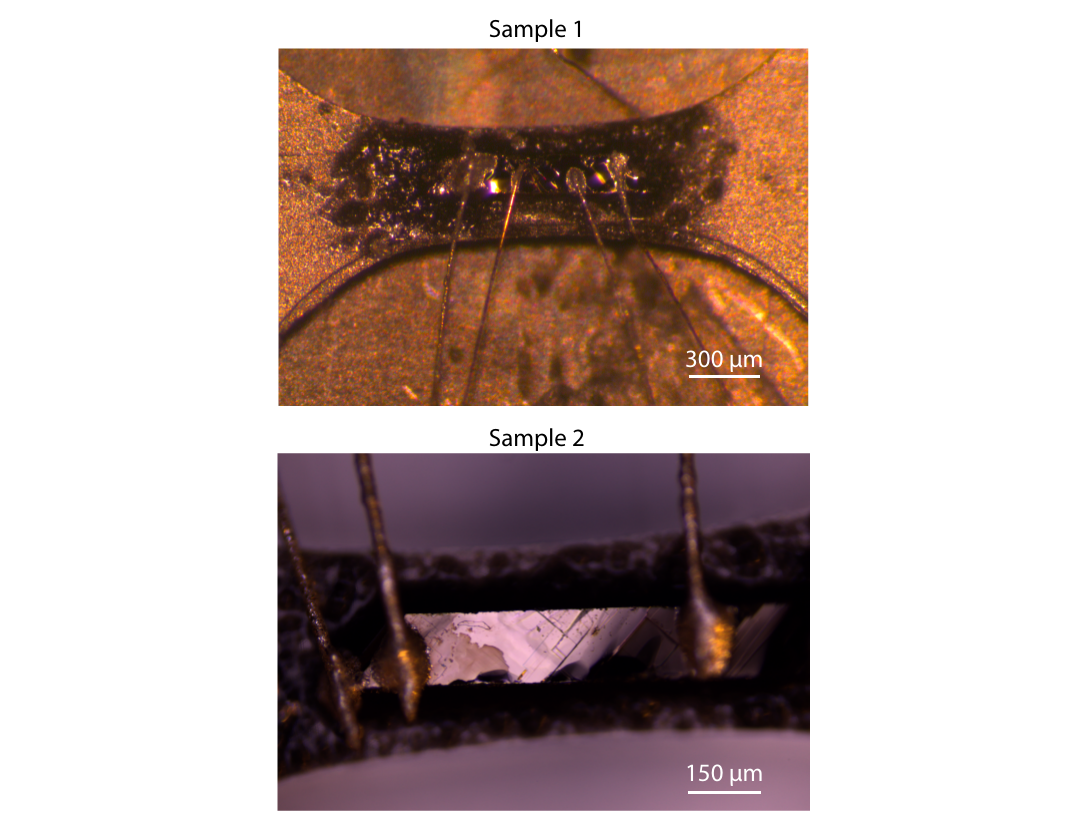}
\end{figure}

\noindent {\bf Fig. S1.} Optical images of Sample 1 (top) and Sample 2 (bottom) after mounting and wiring.\\

\clearpage

\begin{figure}[h]
\centering
\includegraphics[width = 0.5 \columnwidth]{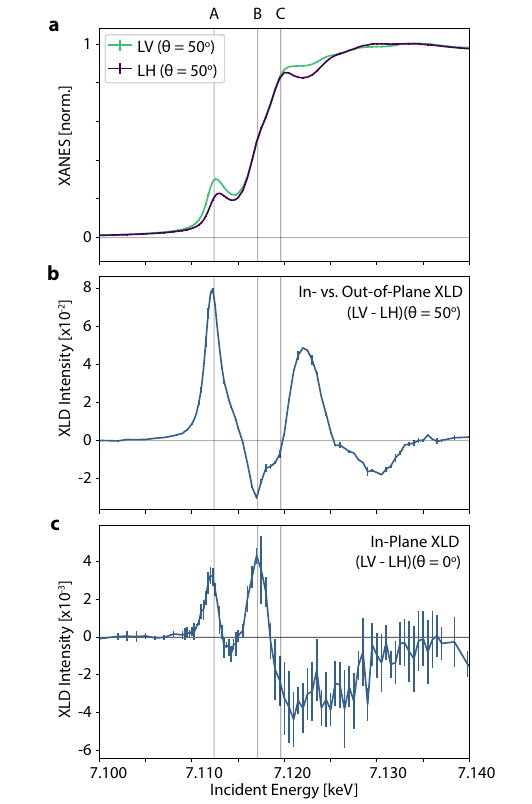}
\end{figure}

\noindent {\bf Fig. S2.} A comparison of the pure in-plane and the  out-of-plane linear dichroism spectra. {\bf a}, XAS spectra in LH/LV polarizations with the beam incident at an angle $\theta = 50^\circ$ with respect to the surface normal within the $\mathbf{b}$-$\mathbf{c}$ plane (see text). {\bf b}, The out-of-plane XLD spectrum corresponding to the polarized XAS spectra in {\bf a}, compared to {\bf c}, the pure in-plane XLD spectrum with LV $\parallel \mathbf{b}$ and LH $\parallel \mathbf{a}$ with the beam at normal incidence along the $\mathbf{c}$ crystallographic direction. Note: the out-of-plane XLD spectrum is expressed in units of $10^{-2}$ while the in-plane are expressed in units of $10^{-3}$. Error bars in all panels are determined from the standard deviation of repeated scans ({\bf a}/{\bf b}, $n = 2$ and {\bf c}, $n = 5$).\\

\clearpage

\begin{figure}[h]
\centering
\includegraphics[width = 0.5 \columnwidth]{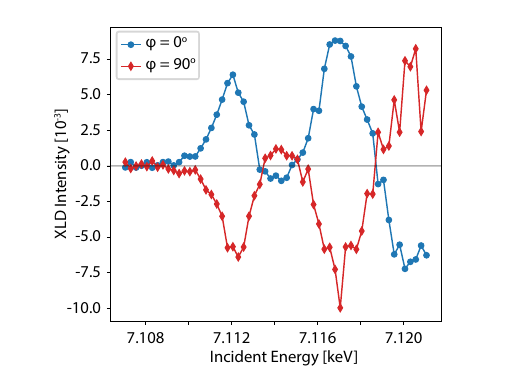}
\end{figure}

\noindent {\bf Fig. S3.} In-plane XLD signal at normal incidence of the fully-detwinned Sample 1 at $T = 50$ K for two different sample orientations rotated around the $\mathbf{c}$ axis by a relative angle $\phi$ of 90 degrees. $\phi = 0^\circ$ corresponds to LV/LH parallel to $\mathbf{b}$/$\mathbf{a}$, respectively, and vice versa for $\phi = 90^\circ$.\\

\clearpage

\begin{figure}[h]
\centering
\includegraphics[width = 0.5 \columnwidth]{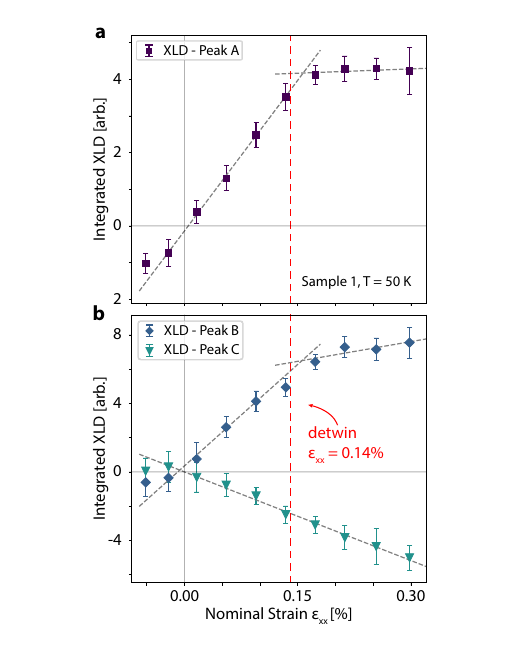}
\end{figure}

\noindent {\bf Fig. S4.} Strain-dependent XLD signal from Sample 1 at $T = 50$ K resolved into the individual XAS features A/B/C (corresponding XAS/XLD spectra are shown in main text Figure 2). {\bf a}, The pre-edge signal saturates for strains larger than the detwinning strain, indicated by the dashed vertical red line at $\epsilon_{xx} \simeq 0.14 \%$. {\bf b}, The higher energy peaks show increasing XLD signal past the detwinning point, with peak C XLD being nearly linear across the whole strain range, while peak B shows a slope change at the detwinning point, but without complete saturation as observed at the pre-edge in {\bf a}. Error bars for all panels are determined from the propagated errors of the XLD spectra (as reported in main text Fig. 2{\bf b}).\\

\clearpage

%FDMNES (S5-S8)

\begin{figure}[h]
\centering
\includegraphics[width = \columnwidth]{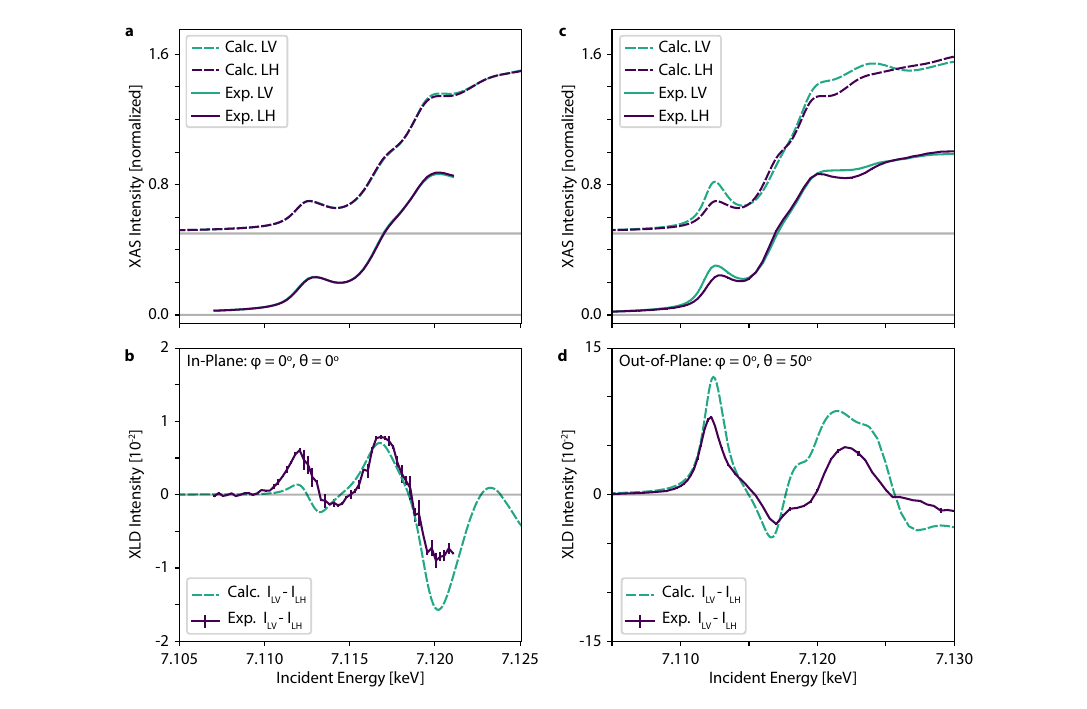}
\end{figure}

\noindent {\bf Fig. S5.} FDMNES calculations of Fe $K$-edge polarized XAS in FeSe. {\bf a}, XAS and {\bf b}, XLD for the in-plane configuration with $\phi = 0^\circ$ and $\theta = 0^\circ$ as described in Fig. S2. {\bf c}, XAS and {\bf d}, XLD in the out-of-plane configuration with $\phi = 0^\circ$ and $\theta = 50^\circ$. In all panels, dashed lines correspond to calculations and solid lines are experimental data in Sample 1, with error bars calculated from the standard deviation of repeated scans ({\bf b}, $n = 5$ and {\bf d}, $n = 2$). The in-plane data is the fully detwinned data at $T = 50$ K from main Figure 2{\bf b}. The out-of-plane data is at $T = 25$ K reproduced from Figure S2. Curves in {\bf a}/{\bf c} are offset for clarity.\\

\clearpage

\begin{figure}[h]
\centering
\includegraphics[width = 0.5 \columnwidth]{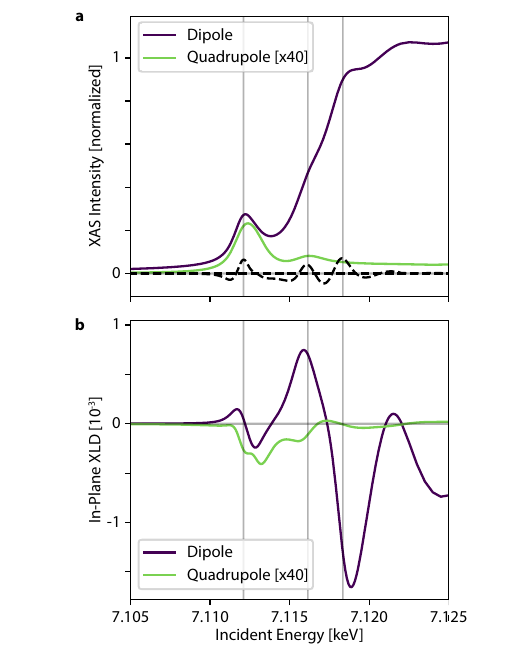}
\end{figure}

\noindent {\bf Fig. S6.} Calculations comparing the transition weights corresponding to dipolar ($1s \to 4p$) and quadrupolar ($1s \to 3d$) transitions. {\bf a}, XAS and {\bf b}, XLD for in-plane configuration resolved into the dipole (purple) and quadrupolar (green) contributions. In both cases, the quadrupolar is multiplied by a factor of 40 for clarity. The dashed line in {\bf a} is the second derivative for identifying the features A/B/C.\\

\clearpage

\begin{figure}[h]
\centering
\includegraphics[width = 0.5 \columnwidth]{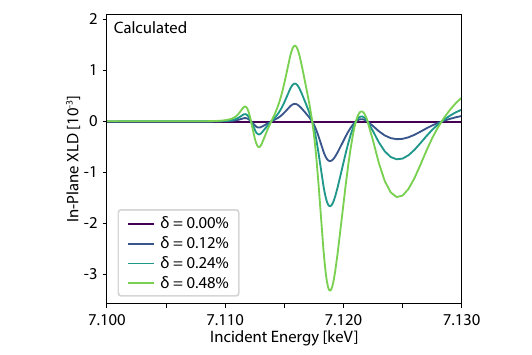}
\end{figure}

\noindent {\bf Fig. S7.} Calculations comparing the expected in-plane XLD as a function of the lattice orthorhombicity ($\delta$). The base temperature orthorhombicity in FeSe is $\delta = 0.24 \%$ as measured in experiment at $T = 10$ K in both Sample 1/2. The structural XLD contributions are all linear in $\delta$.\\

\clearpage

\begin{figure}[h]
\centering
\includegraphics[width = \columnwidth]{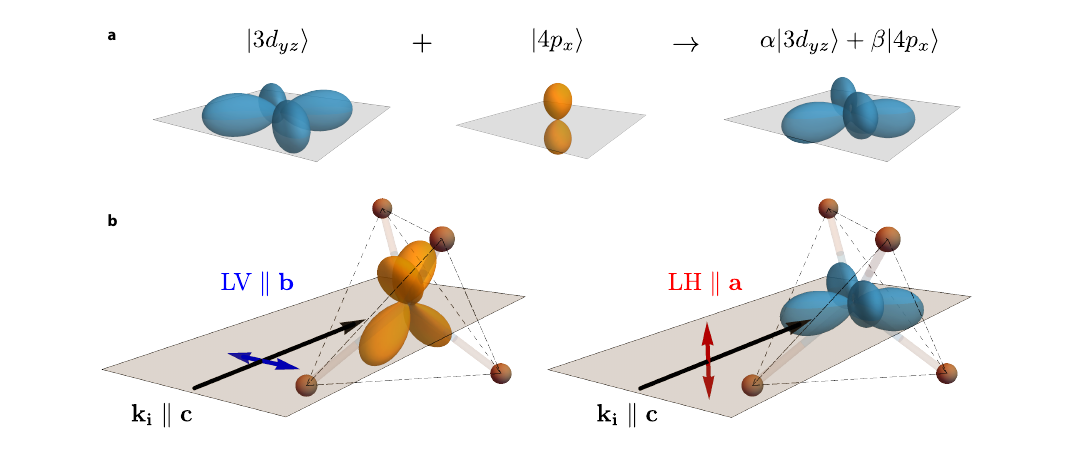}
\end{figure}

\noindent {\bf Fig. S8.} Schematic describing the sensitivity of the Fe $K$ pre-edge measurements to the Fe-$3d$ states. {\bf a}, Schematic showing the on-site Fe $3d$/Fe $4p$ mixing induced by hybridization in the tetrahedral environment. {\bf b}, the $3d_{xz}$ orbital (orange) within the FeSe$_4$ tetrahedral environment is selected by polarization parallel to the $\mathbf{b}$ axis and {\bf c}, the $3d_{yz}$ orbital (blue) is selected by polarization parallel to $\mathbf{a}$. The distortions of the orbitals are exaggerated and for illustration purposes only (for quantitative details, see Refs. \cite{Lee2010} and \cite{Fguieredo2022}).\\

\clearpage

%XRD Sample 1 (S9/S10)

\begin{figure}[h]
\centering
\includegraphics[width = 9cm]{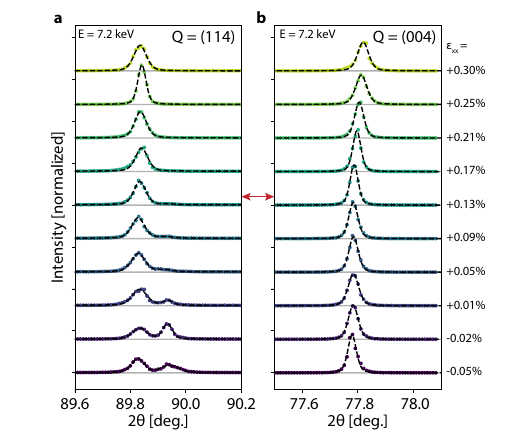}
\end{figure}

\noindent {\bf Fig. S9.} Raw X-ray diffraction scans of the {\bf a}, $(114)$ and {\bf b}, $(004)$ structural peaks used to determine the lattice parameters during the strain loop at $T = 50$ K on Sample 1 as presented in main text Figure 2{\bf e} and 2{\bf f}. Data is acquired with $E = 7.2$ keV incident energy. The corresponding nominal strain values $\epsilon_{xx}$ are shown on the right, and the red arrow indicates the approximate detwinning strain of +$0.14\%$. Dashed black lines are Gaussian fits. Note: the (114) measurement was optimized to track the lattice constants, but not the relative intensity of the A/B domain peaks (see text).\\

\clearpage

\begin{figure}[h]
\centering
\includegraphics[width = \columnwidth]{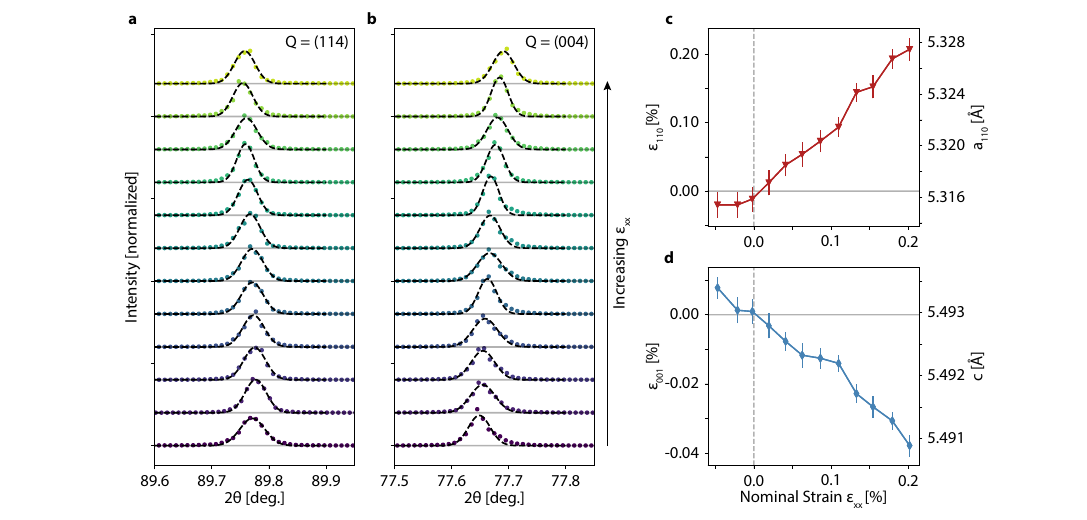}
\end{figure}

\noindent {\bf Fig. S10.} X-ray diffraction data at $T = 120$ K on Sample 1. Data is acquired with incident energy of $E = 7.2$ keV. Raw XRD scans versus increasing $\epsilon_{xx}$ (from bottom to top) for the {\bf a}, (114) and {\bf b}, (004) diffraction peaks with Gaussian fits shown as dashed black lines. The corresponding $[110]$ and $[001]$ lattice parameters (right axes) and associated uniaxial strains (left axes) are shown in {\bf c}/{\bf d} respectively. Error bars in {\bf c}/{\bf d} represent the standard errors from Gaussian fits. Zero-strain was determined by the zero-crossing of simultaneously recorded XLD (shown below in Fig. S17). \\

\clearpage

%Optical Birefringence (S11)

\begin{figure}
\centering
\includegraphics[width = \columnwidth]{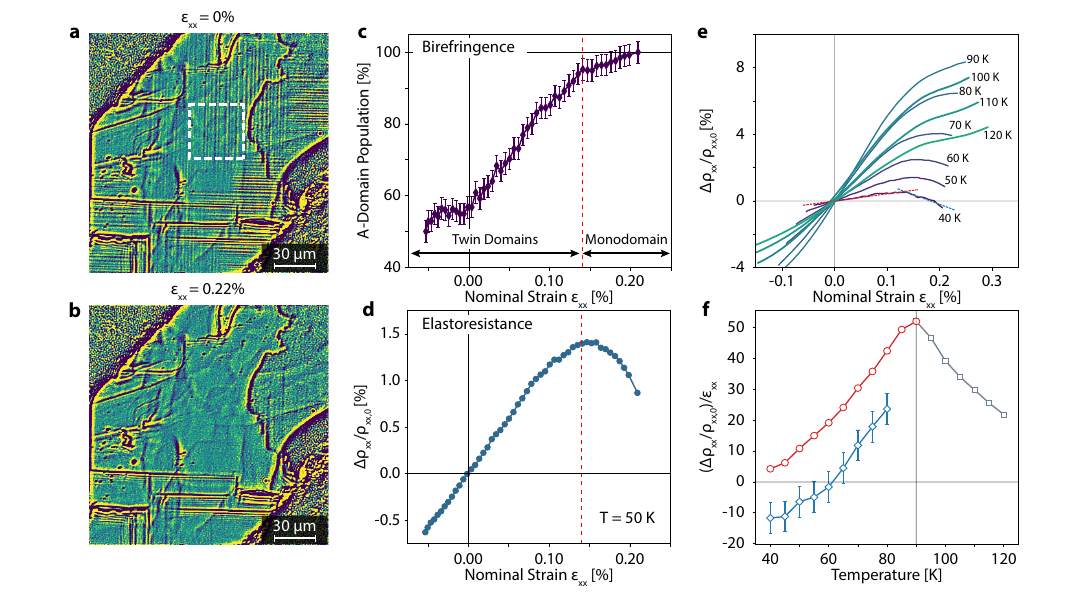}
\end{figure}

\noindent {\bf Fig. S11.}  Optical birefringence and elastoresistivity. {\bf a,b,} Representative wide-field birefringence images of Sample 2 under zero stress ({\bf a}) and under large tension ({\bf b}) at $T = 50$ K. In {\bf a} we define a region of interest (ROI, white box) and estimate the tensile A domain population within. The strain dependence of the A domain population is plotted in {\bf c} and is found to saturate near $\epsilon_{xx}=0.14\%$. Error bars represent the standard deviation of the ROI ($n = 300^2$) in the fully detwinned state ($\epsilon = 0.22\%$). The simultaneous resistivity $\rho_{xx}$ is plotted in {\bf d} and its slope is found to change sign between the twin domain and monodomain strain ranges. In {\bf e} we plot the zero-strain normalized resistivity $\rho_{xx}/\rho_{xx,0}-1$ versus strain at temperatures above and below the nematic transition. In {\bf f} we plot the slope of the resistivity at each temperature over specific strain ranges;  about the zero strain point for $T>T_{s}=90$ K (gray); about the zero strain point for $T< 90$ K (red) to characterize the spontaneous resistivity anisotropy, and at strains beyond the full detwinning point for $T< 80$ K (blue) to characterize the post-detwinning strain-induced elastoresistivity. Error bars in the post-detwinning data represent approximated uncertainty in the slope due to a corresponding uncertainty in the detwinning point.\\

\clearpage

%XLD TDep Sample 2 (S12-S13)

\begin{figure}[h]
\centering
\includegraphics[width = \columnwidth]{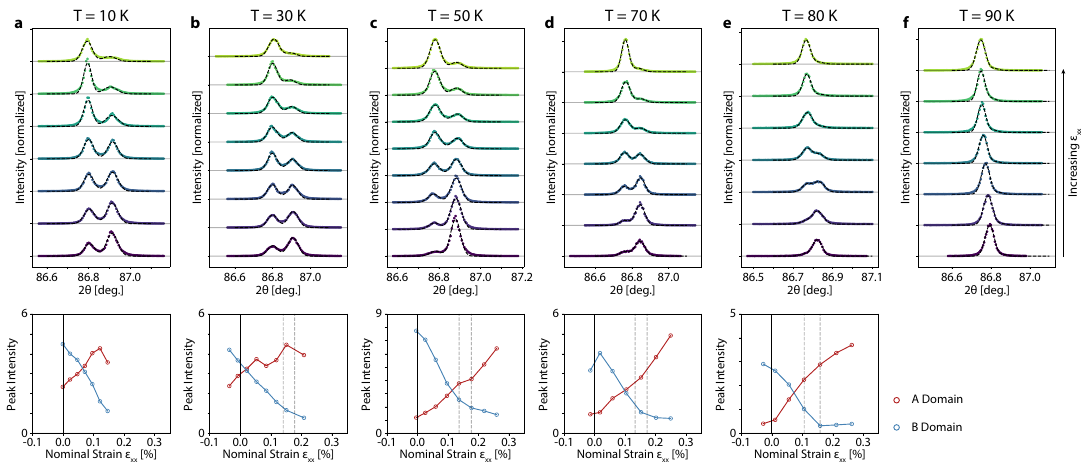}
\end{figure}

\noindent {\bf Fig. S12.} X-ray diffraction data of the $(114)$ Bragg reflection in Sample 2 versus strain and temperature. Data is acquired on separate (but otherwise identically initialized) strain loops immediately after the XLD data strain loop in main text Figure 3. The integrated intensities from Gaussian fits (shown in black dashed line) of the A/B domain twin peaks are shown versus strain in red/blue, respectively, in the bottom row. Vertical grey dashed lines mark the estimated detwinning ranges where the minority domain peaks begin to saturate.\\

\clearpage

\begin{figure}[h]
\centering
\includegraphics[width = \columnwidth]{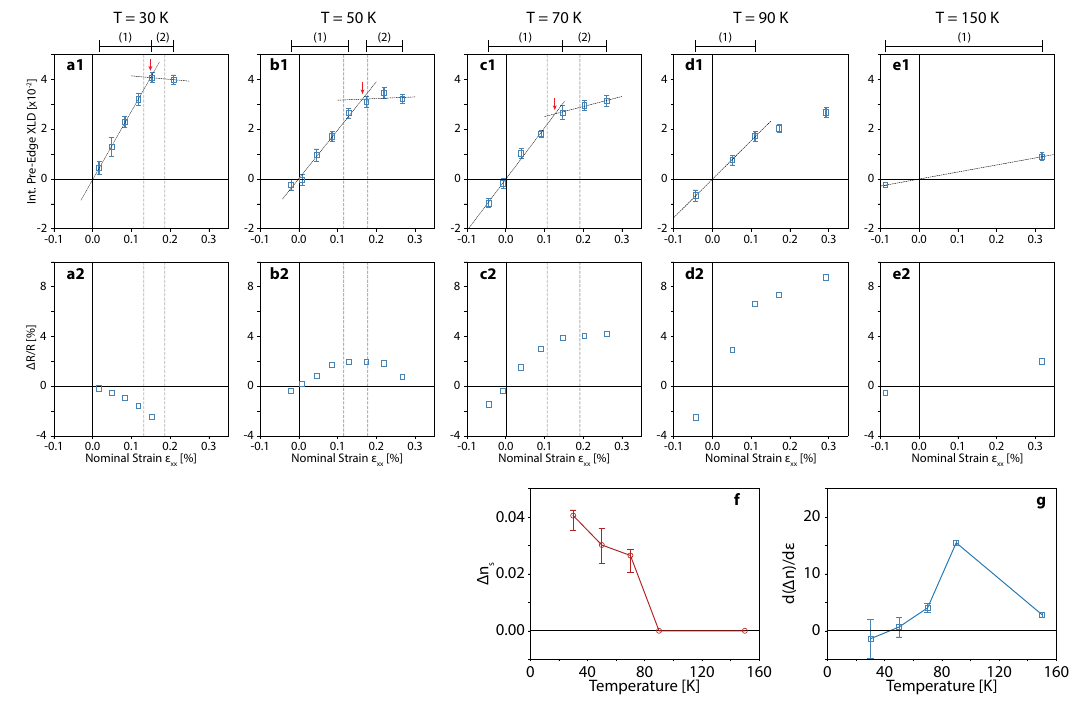}
\end{figure}

\noindent {\bf Fig. S13.}  Additional detail on the detwinning ranges and fit ranges for the strain- and temperature-dependent XLD presented in main text Figure 3. Top row, {\bf a1}-{\bf e1}: integrated pre-edge XLD signal versus nominal strain, with error bars representing the propagated errors from the XLD spectra. Middle row, {\bf a2}-{\bf e2}: simultaneous elastoresistance data. For T = 30/50/70 K ($T<T_s$) the vertical dashed grey lines indicate the approximate detwinning range estimated independently from XRD and birefringence data. Dashed black lines indicate linear fits to the XLD intensity in the two regimes of the data (1) within the twin domain region and (2) in the detwinned regime. The precise data points for each fit are indicated by the ranges (1)/(2) on top of each plot. Red arrows show the intercept of the linear fits. From the approximated detwinning ranges, we extract the spontaneous value of the XLD at detwinning ($\Delta n_s$) shown in {\bf f}. We also estimate the strain-susceptibility of the XLD about zero strain (for $T \geq T_s$, denoted as fit range (1)) and in the monodomain state (for $T < T_s$, denoted as fit range (2)) shown in {\bf g}. Error bars in {\bf f} are determined by the uncertainty in the detwinning point, while errors in {\bf g} are determined by the standard error of the linear fit including the statistical errors of the XLD data.\\

\clearpage

%XLD/XRD Tdep Sample 1 (S14)

\begin{figure}[h]
\centering
\includegraphics[width = \columnwidth]{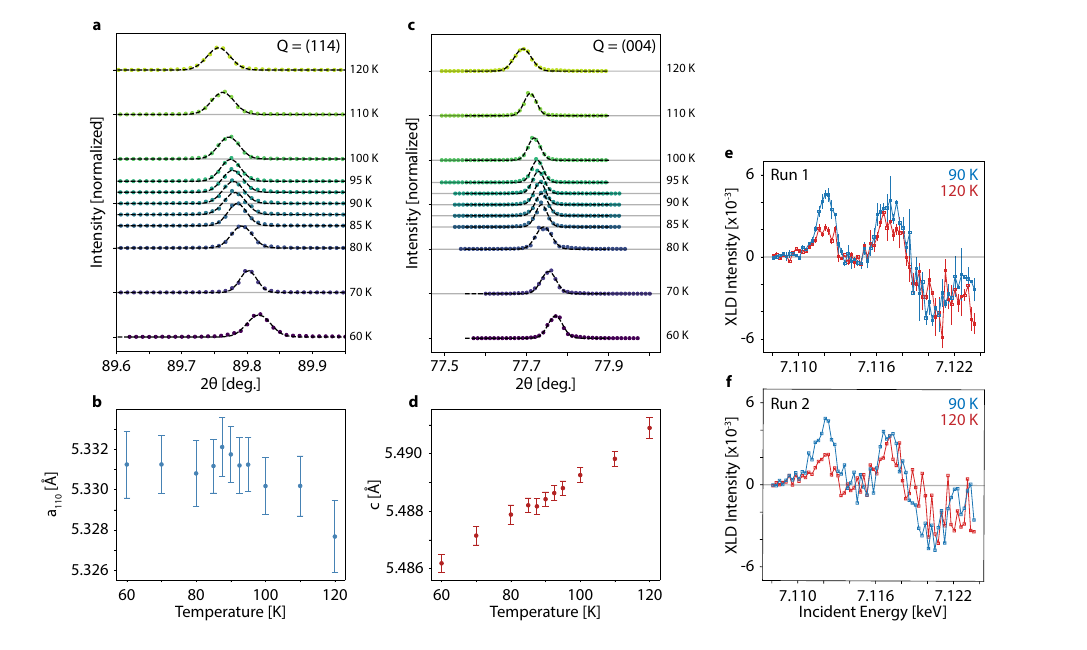}
\end{figure}

\noindent {\bf Fig. S14.} Additional X-ray diffraction and XLD data for the fixed strain temperature dependence data presented in main text Figure 4{\bf a}/{\bf b}. Raw x-ray diffraction scans for the {\bf a}, (114) and {\bf c}, (004) Bragg peaks under fixed tension conditions. Strain was initialized at $T = 120$ K and temperature was monotonically decreased to $60$ K. These data are used to extract the $T$-dependent {\bf b}, $[110]$ and {\bf d}, $[001]$ lattice parameters. XLD data were taken on two separate but identically prepared temperature dependence runs. Representative data at $T_s = 90$ K (blue) and $120$ K (red) are compared in the respective runs in panels {\bf e} and {\bf f}, showing roughly temperature independent high-energy XLD features B/C and a large temperature dependence of the pre-edge feature that is reproduced on both runs. XLD (standard deviation $n = 2$) and lattice parameter error bars are the same as defined in the main text.\\

\clearpage

%XLD vs Resistance (S15-S18)

\begin{figure}[h]
\centering
\includegraphics[width = \columnwidth]{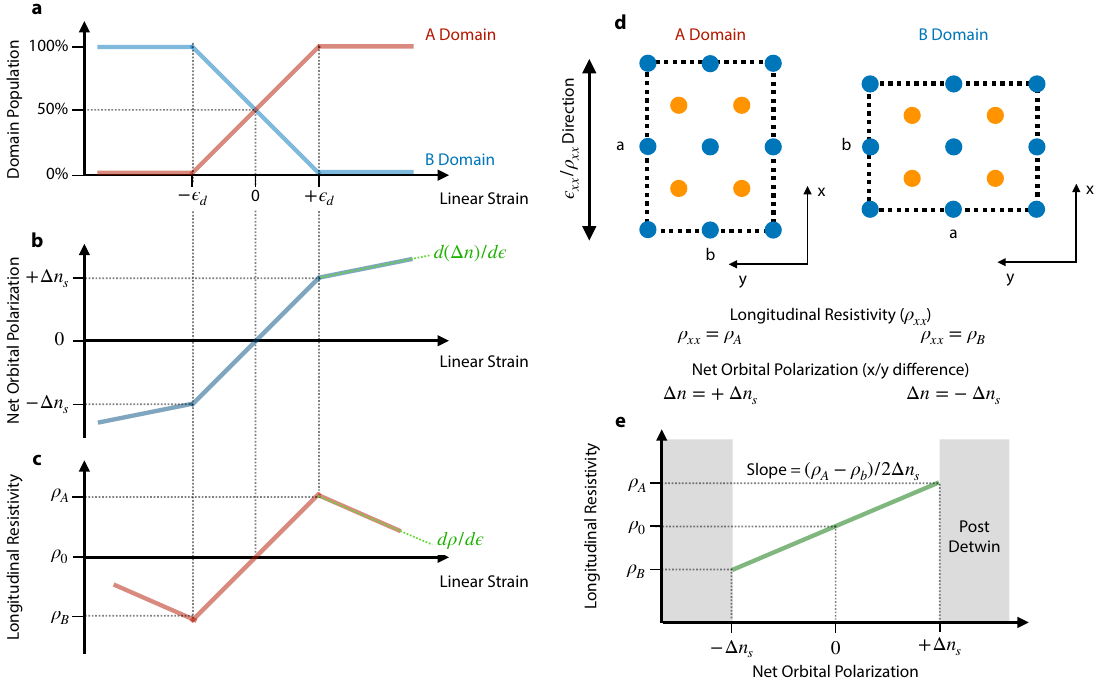}
\end{figure}

\noindent {\bf Fig. S15.} Schematic of the framework used to interpret the nematic-phase strain-dependent data throughout the paper. {\bf a}, The relative volume fraction of the A and B domain populations, with full detwinning strains defined as $\pm \epsilon_d$. {\bf b} The net orbital polarization measured in a fixed XLD geometry, with spontaneous values of $\pm \Delta n_s$ at $\pm \epsilon_d$ and beyond-detwinning strain susceptibility $d(\Delta n)/d\epsilon$. {\bf c} The longitudinal resistivity, with spontaneous values of $\rho_A$ and $\rho_B$ at $\pm \epsilon_d$ and beyond-detwinning strain susceptibility $d\rho/d\epsilon$ (when further normalized by $\rho_0$, this is also known as the elastoresistivity). {\bf d}, Schematic of the twin domains in relation to a fixed reference frame ($x$/$y$) and measurement geometry, with defined spontaneous quantities for the orbital polarization $\Delta n_s$ and resistivity $\rho_A$/$\rho_B$. {\bf e}, Schematic of the behavior of the longitudinal resistivity versus the net orbital polarization within the twin domain region, demonstrating the expected linearity and its relation to the spontaneous resistivity anisotropy $\Delta \rho_{s} = \rho_A - \rho_B$ and the spontaneous orbital polarization $\Delta n_{s}= n_{x} - n_{y}$.\\

\clearpage

\begin{figure}[h]
\centering
\includegraphics[width = \columnwidth]{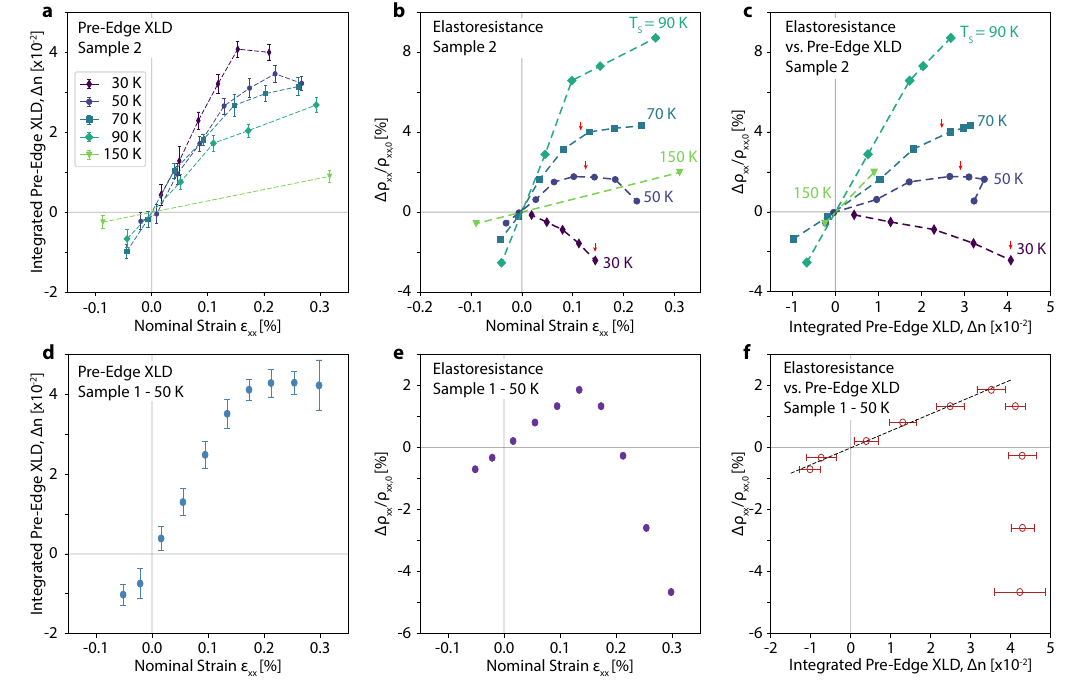}
\end{figure}

\noindent {\bf Fig. S16.} Elastoresistivity versus XLD data in Samples 1 and 2. {\bf a}, XLD and {\bf b}, elastoresistance versus strain and temperature from Sample 2. The XLD data is reproduced from main text Fig. 3{\bf c}/{\bf d} in the main text and the elastoresistance data is collected simultaneously to the XLD data. {\bf c}, The same data in {\bf a}/{\bf b} plotted as elastoresistance versus XLD. Red arrows denote the approximate detwinning point. The analogous data for Sample 1 at $T = 50$ K is shown in panels {\bf d}-{\bf f}, which is reproduced from main text Fig. 2{\bf d}/{\bf g} and Fig. 4{\bf c}(inset) for comparison between the samples. The slopes of the elastresistance versus XLD within the detwinning regime are used for main text Fig. 4{\bf c} summarizing the orbital-transport correspondence. Integrated XLD error bars in all panels are as defined in the main text.\\

\clearpage

\begin{figure}[h]
\centering
\includegraphics[width = \columnwidth]{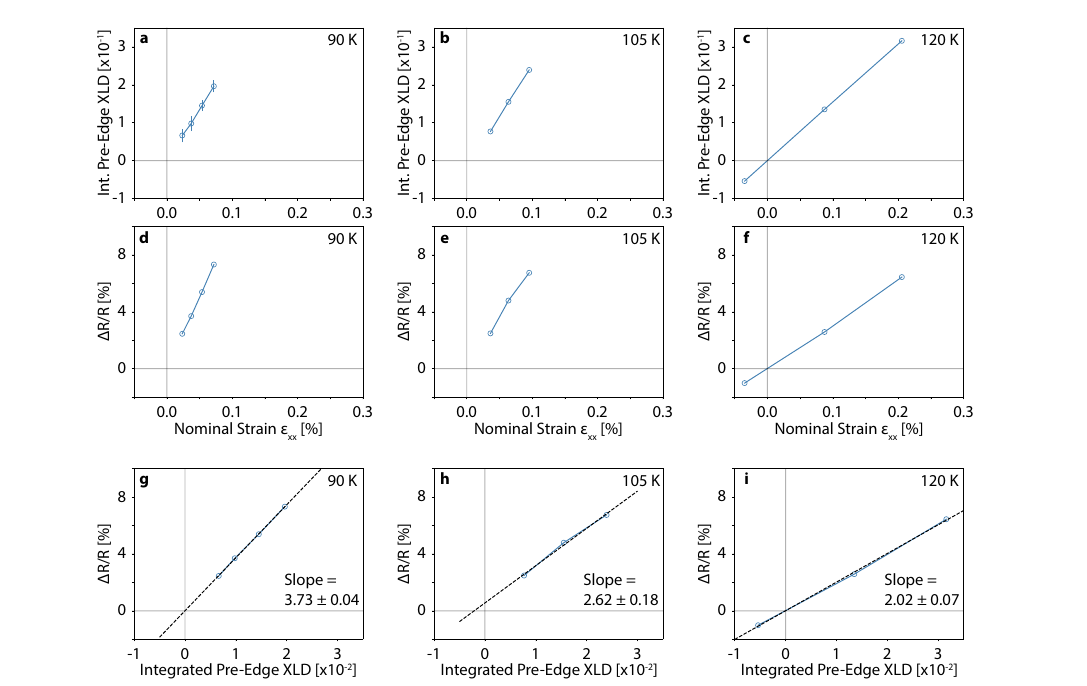}
\end{figure}

\noindent {\bf Fig. S17.} Pre-edge XLD (top row, {\bf a}-{\bf c}) and elastoresistance ($\Delta R/R$, middle row, {\bf d}-{\bf f}) over short, monotonically-increasing nominal strain $\epsilon_{xx}$ loops for $T =$ 90/105/120 K from left to right on Sample 1. The data in {\bf c}/{\bf f} at $T = 120$ K corresponds to the same strain loop for the XRD data presented in Fig. S10 above. These data are used to plot the elastoresistance versus XLD slopes in {\bf g}-{\bf i} for each temperature, with linear fits shown as dashed black lines. The calculated slopes are shown on the respective figures, which are used for the correspondent Sample 1 data in the summary plot in main text Figure 4{\bf c}. Integrated XLD error bars in {\bf a} are as defined in the main text. Statistical errors for $T = 90$ K and $T = 105$ K could not be estimated as only a single XLD spectrum was recorded for these temperatures.\\

\clearpage

\begin{figure}[h]
\centering
\includegraphics[width = \columnwidth]{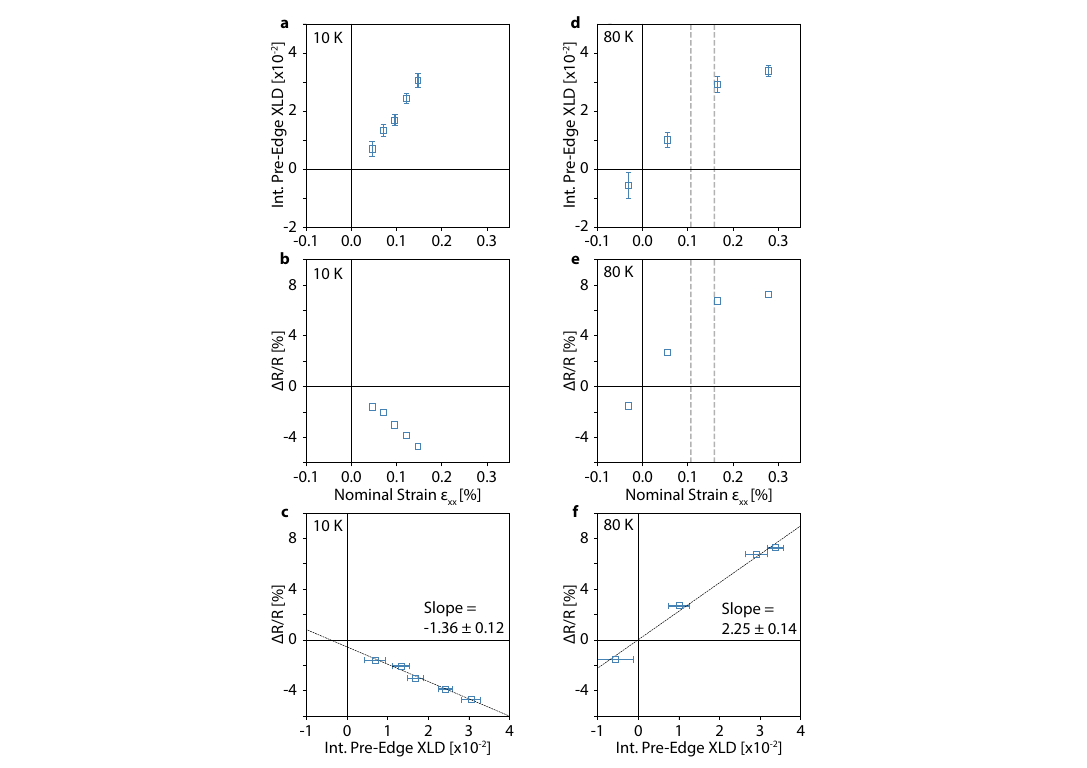}
\end{figure}

\noindent {\bf Fig. S18.}  Additional XLD and elastoresistance data on Sample 2 collected using the same procedures as the data in main text Figure 3 and on the same cryostat cooldown at {\bf a}-{\bf c}, 10 K and {\bf d}-{\bf f}, 80 K. Top row, integrated pre-edge XLD; middle row, simultaneous elastoresistance data; and bottom row, Elastoresistance versus XLD plot. In {\bf d}/{\bf e}, vertical lines indicate the approximate detwinning point estimated from the XRD data in Figure S12. Dashed lines in {\bf c}/{\bf f} are linear fits utilizing all data points with the slopes (and standard errors) indicated. These data are used for the summary plot in main text Figure 4{\bf c}.\\

%\bibliographystyle{naturemag}
%\bibliography{FeSe_XLD_references_corrected_20230101}